\documentclass[longauth]{aa} % for the long lists of affiliations 
\usepackage{graphicx}
\usepackage{graphics}
\usepackage{amssymb}
\usepackage{txfonts}
\usepackage{mathrsfs}
\usepackage{array}
\usepackage{float}
\usepackage{ulem}
\usepackage{longtable,lscape} 
\usepackage{array} 
\usepackage{tabularx}
\usepackage{amsmath}
\usepackage{wrapfig}
\usepackage{natbib}
\bibpunct{(}{)}{;}{a}{}{,}   % to follow the A&A style
\begin{document}
\title{The VISTA Variables  in the V\'ia L\'actea  eXtended (VVVX) ESO public
  survey:    Completion     of    the    observations    and
  legacy\thanks{Based  on  observations  taken within  the  ESO  VISTA
    Public  Survey   VVV  and  VVVX,  Programmes   ID  179.B-2002  and
    198.B-2004, respectively.}}

\author{R.~K.~Saito\,$^{1}$
\and M.~Hempel\,$^{2,3}$
\and J.~Alonso-Garc\'ia\,$^{4,5}$
\and P.~W.~Lucas\,$^{6}$
\and D.~Minniti\,$^{2,7,1}$
\and S.~Alonso\,$^{8}$
\and L.~Baravalle\,$^{9,10}$
\and J.~Borissova\,$^{11,12}$
\and C.~Caceres\,$^{2}$
\and A.~N.~Chen\'e\,$^{13}$
% ...
}
\offprints{to: roberto.saito@ufsc.br} 

\institute{    
$^{1}$ Departamento de F\'isica, Universidade Federal de Santa Catarina,
  Trindade 88040-900, Florian\'opolis, Brazil\\
$^{2}$ Instituto  de  Astrof\'isica,   Dep.  de  Ciencias  F{\'i}sicas,
  Facultad  de   Ciencias  Exactas,  Universidad  Andres   Bello,  Av.
  Fern\'andez Concha 700, Santiago, Chile\\
$^{3}$  Max-Planck  Institute  for  Astronomy,  Königstuhl  17,  69117
  Heidelberg, Germany\\
$^{4}$ Centro de  Astronom\'{i}a (CITEVA), Universidad  de Antofagasta,
  Av. Angamos 601, Antofagasta, Chile\\
$^{5}$ Millennium  Institute of Astrophysics (MAS),  Nuncio Monse\~nor
  Sotero Sanz 100, Of. 104, Providencia, Santiago, Chile\\
$^{6}$ Centre for  Astrophysics Research, University  of Hertfordshire,
  College Lane, Hatfield AL10 9A, UK\\
$^{7}$ Vatican  Observatory, Specola Vaticana, V-00120,  Vatican City,
  Vatican City State\\
$^{8}$ Departamento  de Geof\'isica y Astronom\'ia,  CONICET, Facultad
  de Ciencias Exactas, F\'isicas  y Naturales, Universidad Nacional de
  San Juan, Av.  Ignacio de la Roza 590 (O),  J5402DCS, Rivadavia, San
  Juan, Argentina\\
$^{9}$   Instituto   de    Astronom\'ia   Te\'orica   y   Experimental,
  (IATE-CONICET), Laprida 854, X5000BGR, C\'ordoba, Argentina\\
$^{10}$ Observatorio  Astron\'omico de C\'ordoba,  Universidad Nacional
  de C\'ordoba, Laprida 854, X5000BGR, C\'ordoba, Argentina\\
$^{11}$   Instituto  de   F\'isica  y   Astronom\'ia,  Universidad   de
  Valpara\'iso, ave. Gran Breta\~na, 1111, Casilla 5030, Valpara\'iso,
  Chile\\
$^{12}$ Millennium Institute of  Astrophysics, Nuncio Monse\~nor Sotero
  Sanz 100, Of. 104, Providencia, Santiago, Chile\\
$^{13}$ Gemini  Observatory, Northern  Operations Center,  670 A'ohoku
  Place, Hilo, HI 96720, USA\\
$^{14}$ Wide-Field Astronomy Unit, Institute for Astronomy, University
  of Edinburgh, Royal Observatory,  Blackford Hill, Edinburgh EH9 3HJ,
  UK\\
$^{15}$  European  Southern  Observatory, Alonso  de  C\'ordova  3107,
  Casilla 19001, Vitacura, Santiago, Chile\\
$^{16}$  European  Southern   Observatory,  Karl  Schwarzschildstr  2,
  D-85748 Garching bei München, Germany\\
$^{17}$  INAF  –  Osservatorio   Astronomico  di  Capodimonte,  Salita
  Moiariello 16, 80131 Napoli, Italy\\
$^{18}$ Mount Saint Vincent University, Halifax B3M2J6, Canada\\
$^{19}$ INAF -  Osservatorio Astronomico di Roma, Via  di Frascati 33,
  I-00078, Monte Porzio Catone, Roma, Italy\\
$^{20}$  Department  of  Physics  and  Astronomy,  University  College
  London, London WC1E 6BT, UK\\
$^{21}$ Institute of Astronomy, University  of Cambridge, Madingley Rd.,
  Cambridge CB3 0HA, UK\\
$^{22}$ Departamento de F{\'i}sica,  Universidade Estadual de Feira de
  Santana  (UEFS), Av.  Transnordestina, S/N,  CEP 44036-900  Feira de
  Santana, BA, Brazil\\ 
$^{23}$ Gemini Observatory/NSF's NOIRLab, Casilla 603, La Serena, Chile\\
$^{24}$ Departamento de Astronom\'ia, Universidad de La Serena, Av. Juan
  Cisternas 1200 Norte, La Serena, Chile\\
$^{25}$ Universidade de S\~ao Paulo,  IAG, Rua do Mat\~ao 1226, Cidade
  Universit\'aria, S\~ao Paulo 05508-090, Brazil\\
$^{26}$ Fundaci\'on Chilena de Astronom\'ia, El Vergel 2252, Santiago,
  Chile\\
$^{27}$ Istituto Nazionale di Astrofisica, Osservatorio Astronomico di
  Padova, Vicolo dell'Osservatorio 5, Padova, IT-35122, Italy\\
$^{28}$  Space Telescope  Science  Institute, 3700  San Martin  Drive,
  Baltimore, MD, 21218, USA\\
$^{29}$ Department of Physics, University of Wisconsin-Whitewater, 800
  West Main Street, Whitewater, WI 53190, USA\\
$^{30}$  Departamento de  Astronomia,  Instituto  de F\'isica,  UFRGS,
  Av. Bento Gon\c{c}alves 9500, Porto Alegre, RS, Brazil\\
$^{31}$  Instituto   de  Ci\^encias   Naturais,  Humanas   e  Sociais,
  Universidade  Federal  de  Mato Grosso,  Cidade  Jardim,  78550-728,
  Sinop, Brazil\\
$^{32}$ Vatican Observatory, VORG,  Steward Observatory, 933 N. Cherry
  Avenue, Tucson, AZ, USA\\
$^{33}$ Gerencia  De Vinculaci\'on Tecnol\'ogica,  Comisi\'on Nacional
  de   Actividades  Espaciales   (GVT-CONAE),   Falda  del   Ca\~nete,
  C\'ordoba, Argentina\\
$^{34}$  Col\'egio Militar  de Porto  Alegre, Minist\'erio  da Defesa,
  Ex\'ercito  Brasileiro, Av.  Jos\'e  Bonif\'acio  363, Porto  Alegre
  90040-130, RS, Brazil\\
$^{35}$ Instituto de Alta  Investigaci\'on, Universidad de Tarapac\'a,
  Casilla 7D, Arica, Chile\\
$^{36}$ Instituto de Astrof\'{i}sica,Pontificia Universidad
  Cat\'{o}lica de Chile, Av. Vicu\~{n}a Mackenna 4860, 7820436 Macul,
  Santiago, Chile\\
$^{37}$   Centro  de   Astro-Ingenier\'{i}a,  Pontificia   Universidad
  Cat\'{o}lica de Chile, Av.  Vicu\~{n}a Mackenna 4860, 7820436 Macul,
  Santiago, Chile\\
$^{38}$ Consejo Nacional de Investigaciones Cient\'ificas y T\'ecnicas
  (CONICET),  Godoy  Cruz  2290,  Ciudad  Autónoma  de  Buenos  Aires,
  C1425FQB, Argentina\\ 
$^{39}$   Department  of   Physics  and   Astronomy,  Seoul   National
  University, Seoul 08826, Republic of  Korea, 2 Research Institute of
  Basic Sciences, Seoul National  University, Seoul 08826, Republic of
  Korea\\
$^{40}$  Departamento   de  Tecnolog\'ias  Industriales,   Faculty  of
  Engineering, Universidad de Talca, Merced 437, Curic\'o, Chile\\
$^{41}$  Departamento de  F\'isica  de la  Tierra  y Astrof\'isica  \&
  IPARCOS-UCM (Instituto de F\'isica de Part\'iculas y del Cosmos de la
  UCM),  Facultad de  Ciencias F\'isicas,  Universidad Complutense  de
  Madrid, 28040 Madrid, Spain\\
$^{42}$  Centro de  Astrobiología  (CAB), CSIC-INTA,  Camino Bajo  del
  Castillo s/n, E-28692, Villanueva de la Ca\~nada, Madrid, Spain\\
$^{43}$   Facultad   de   Matem\'atica,   Astronom\'ia,   F\'isica   y
  Computaci\'on, Universidad  Nacional de C\'ordoba  (UNC), C\'ordoba,
  Argentina\\
$^{44}$ University of Central Lancashire, Preston, PR1 2HE, UK\\
$^{45}$ Instituto de Ciencias Astron\'omicas, de la Tierra y del Espacio
  (ICATE, CONICET), C.C. 467, 5400, San Juan, Argentina\\
$^{46}$  Astronomy Unit,  School  of Physical  and Chemical  Sciences,
  Queen Mary University of London, Mile End Road, London, E1 4NS, UK\\
$^{47}$ Instituto  de Astronom\'ia, Universidad Cat\'olica  del Norte,
  Av. Angamos 0610, Antofagasta, Chile\\
$^{48}$ Department of Astronomy, Yale University, 219 Prospect Street,
  New Haven, CT 06511, USA\\
$^{49}$ Instituto de Astronom\'ia  y Ciencias Planetarias, Universidad
  de Atacama, Copayapu 485, Copiap\'o, Chile\\
$^{50}$ Laboratorio Nacional de Astrof\'isica LNA/MCTI, 37504-364
  Itajub\'a, MG, Brazil\\
$^{51}$  Centre  for Astrophysics  and  Planetary  Science, School  of
  Physics and Astronomy, University of Kent, Canterbury CT2 7NH, UK\\
$^{52}$  Departamento de  Astronomia,  Casilla  160-C, Universidad  de
  Concepcion, Chile\\
$^{53}$ Instituto  Multidisciplinario de Investigaci\'on  y Postgrado,
  Universidad de La Serena, Ra\'ul Bitr\'an 1305, La Serena, Chile\\
$^{54}$ Max-Planck-Institut f\"ur Ex. Physik, Giessenbachstrasse, 85748,
  Garching, Germany\\
$^{55}$ UK  Astronomy Technology Centre, Royal  Observatory Edinburgh,
  Blackford Hill, EH9 3HJ, Edinburgh, United Kingdom\\
$^{56}$ Universit\'e C\^ote d'Azur,  Observatoire de la C\^ote d'Azur,
  CNRS,  Laboratoire Lagrange,  Blvd de  l’Observatoire, 06304,  Nice,
  France\\
$^{57}$  Centro  Internacional  Franco  Argentino de  Ciencias  de  la
  Informaci\'on  y   de  Sistemas  (CIFASIS,   CONICET–UNR),  Rosario,
  Argentina\\
$^{58}$ Dipartimento di Fisica,  Universit\`a di Ferrara, Via Giuseppe
  Saragat 1, Ferrara I-44122, Italy\\
$^{59}$ Western Sydney University, Kingswood campus, NSW, Australia\\
$^{60}$  Centro de  Astrof\'isica da  Universidade do  Porto, Rua  das
  Estrelas, s/n, 4150-762, Porto, Portugal\\ 
$^{61}$ Miranda House, University of Delhi, India\\
$^{62}$ Inter University centre  for Astronomy and Astrophysics, Pune,
  India\\
$^{63}$  Universidade  Federal do  Pampa  Br  472  -  Km 585,  CP  118
  Uruguaiana, RS, Brazil\\
$^{64}$ Department of Physics \&  Astronomy, Texas Tech University, Box
  41051, Lubbock TX 79409-1051, USA\\
$^{65}$ Instituto de Astrof\'isica de Canarias, Spain\\
$^{66}$  Departamento  de  Astrof\'isica, Universidad  de  La  Laguna,
  Spain\\
$^{67}$ Istituto Nazionale di Astrofisica, Osservatorio di Astrofisica
  e  Scienza  dello  Spazio  di Bologna,  Via  Gobetti  101,  I-40129,
  Bologna, Italy\\
$^{68}$  Department  of Astronomy,  Graduate  School  of Science,  The
  University of Tokyo, 7-3-1 Hongo, Bunkyo-ku, Tokyo 113-0033, Japan\\
$^{69}$ Jodrell  Bank Centre  for Astrophysics, Department  of Physics
  and Astronomy, The University of Manchester, Oxford Road, Manchester
  M13 9PL, UK\\
$^{70}$ Departamento de Astronom\'ia,  Universidad de Chile, Camino El
  Observatorio 1515, Las Condes, Chile\\
$^{71}$ Association of Universities  for Research in Astronomy (AURA),
  Chile\\
$^{72}$  Grupo  de  Astrof\'isica  Extragal\'actica-IANIGLA,  CONICET,
Universidad   Nacional  de   Cuyo  (UNCuyo),   Gobierno  de   Mendoza,
Argentina\\
$^{73}$ Department of Physics and Astronomy, Johns Hopkins University,
Baltimore, MD 21218, USA\\
$^{74}$  Cl\'inica  Universidad  de   los  Andes,  Chile,  Direcci\'on
Comercial\\
$^{75}$  SOAR  Telescope/NSF's  NOIRLab,  Avda  Juan  Cisternas  1500,
1700000, La Serena, Chile\\
$^{76}$ Centro de Docencia Superior en Ciencias B\'asicas, Universidad
Austral de Chile, Los Pinos s/n, Puerto Montt, Chile\\
$^{77}$ Centre for Basic Space  Science, University of Nigeria, 410101
Nsukka, Nigeria\\
$^{78}$ Departamento de Matem\'atica, Universidad de Atacama, Copayapu
485, Copiap\'o, Chile\\
$^{79}$ NSF NOIRLab/Vera C. Rubin Observatory, Casilla 603, La Serena,
Chile\\
$^{80}$  Instituto  de  Astronom\'ia  y F\'isica  del  Espacio  (IAFE,
CONICET-UBA), C1428ZAA, Ciudad Autónoma de Buenos Aires, Argentina\\
$^{81}$ Instituto de Estudios Astrof\'isicos, Facultad de Ingenier\'ia
y Ciencias, Universidad Diego Portales, Av. Ej\'ercito Libertador 441,
Santiago, Chile\\
$^{82}$    Max-Planck-Institut   f\"ur    extraterrestrische   Physik,
Gie{\ss}enbachstra{\ss}e 1, 85748 Garching, Germany\\
$^{83}$ Th\"uringer Landessternwarte,  Sternwarte 5, 07778 Tautenburg,
Germany\\
$^{84}$  Department of  Physics,  University of  Warwick, Gibbet  Hill
Road, Coventry CV4 7AL, UK\\
$^{85}$  Astrobiology Center,  2-21-1 Osawa,  Mitaka, Tokyo  181-8588,
Japan\\
$^{86}$  National Astronomical  Observatory  of  Japan, 2-21-1  Osawa,
Mitaka, Tokyo 181-8588, Japan\\
$^{87}$  Departamento de  F\'isica, FACI,  Universidad de  Tarapac\'a,
Casilla 7D, Arica, Chile\\
$^{88}$ Museo Interactivo de la Astronom\'ia, Centro Interactivo de la
Conocimientos, Avenida Punta Arenas 6711, La Granja, Chile\\
$^{89}$  North  Optics, Crist\'obal  Col\'on  \#  352 oficina  514,  La
Serena\\
$^{90}$  School  of  Mathematical  and  Physical  Sciences,  Macquarie
University, Sydney, NSW 2109, Australia\\
}
  
   \date{Received ; Accepted }


\abstract
%  context (optional) 
{The ESO  public survey  VISTA Variables in  the V\'ia  L\'actea (VVV)
  surveyed the inner Galactic bulge and the adjacent southern Galactic
  disk from  $2009-2015$. Upon  its conclusion, the  complementary VVV
  eXtended (VVVX)  survey has  expanded both the  temporal as  well as
  spatial coverage of the original VVV area, widening it from $562$ to
  $1700$~sq.~deg., as well as providing additional epochs in $JHK_{\rm
    s}$ filters from $2016-2023$.}
%  aims
{With the completion of VVVX observations during the first semester of
  2023, we present here the  observing strategy, a description of data
  quality and access, and the legacy of VVVX.}
% method
{VVVX took $\sim$\,$2000$~hours, covering about  4\% of the sky in the
  bulge and southern disk. VVVX covered  most of the gaps left between
  the VVV and the VISTA Hemisphere Survey (VHS) areas and extended the
  VVV  time  baseline  in  the   obscured  regions  affected  by  high
  extinction and hence hidden from optical observations.}
% results
{VVVX   provides  a   deep   $JHK_{\rm  s}$   catalogue  of   $\gtrsim
  1.5\times10^9$  point  sources,  as   well  as  a  $K_{\rm  s}$ band
  catalogue of  $\sim$\,$10^7$ variable  sources. Within  the existing
  VVV area, we produced a $5D$ map of the surveyed region by combining
  positions, distances, and proper motions of well-understood distance
  indicators  such  as   red  clump  stars,  RR   Lyrae,  and  Cepheid
  variables.}
% conclusion
{In March 2023  we successfully finished the  VVVX survey observations
  that started in 2016, an  accomplishment for ESO Paranal Observatory
  upon  4200  hours  of   observations  for  VVV+VVVX.   The  VVV+VVVX
  catalogues  complement those  from  the {\it  Gaia}  mission at  low
  Galactic latitudes and provide spectroscopic targets for the
  forthcoming ESO high-multiplex spectrographs MOONS and 4MOST.}

\authorrunning{Saito et al.}
\titlerunning{VVVX completion \& legacy} 
\maketitle
%
%________________________________________________________________

\section{Introduction}
\label{sec:intro}

Despite large-scale  optical surveys  over many decades,  the internal
structure of the  inner regions of the Milky Way  (MW) and the details
of its formation and evolution were poorly understood. The main reason
is  the severe  and non-uniform  interstellar extinction  and crowding
towards the MW  bulge and inner disk,  which complicates observations,
especially at  the optical wavelengths.   These inner regions  are the
most  complex of  our  Galaxy  to study,  with  a  mixture of  stellar
populations  from the  inner disk,  bulge, and  halo, which  exhibit a
variety of physical properties.

\begin{figure*}
\begin{centering}
\includegraphics[scale=0.94]{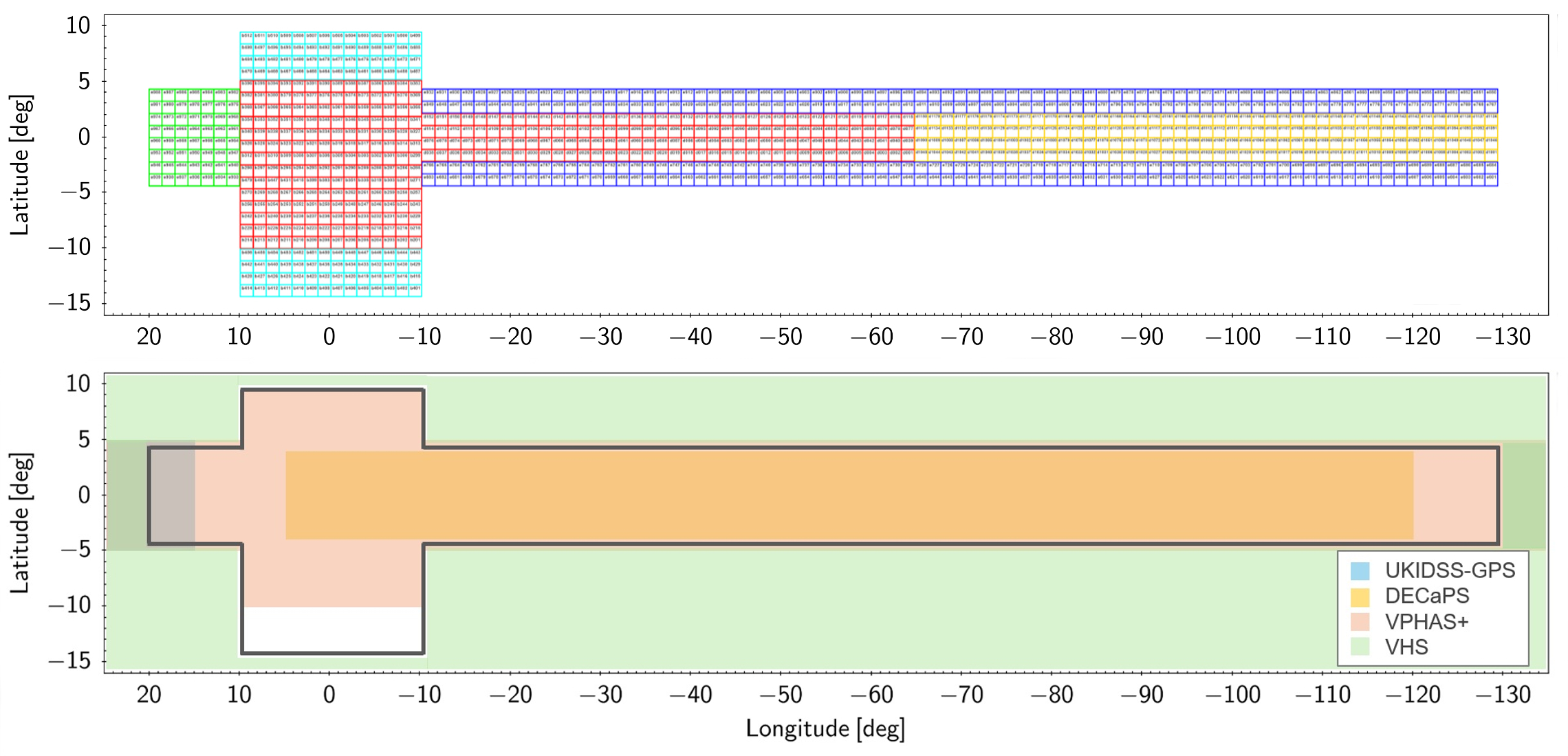}
\caption{VVV+VVVX survey coverage of the  MW bulge and southern plane.
  Top:  The surveyed  area  shown Galactic  coordinates.  Our  near-IR
  survey covers  $\sim$\,1700 sq.  deg.   in total, and  the different
  regions are  colour-coded according  to the location,  baseline, and
  number of observations (see  Section~\ref{sec:strategy}).  In red is
  the original  VVV Bulge and  VVV Disk, with observations  from years
  2010  and  2016   using  the  $ZYJHK_{\rm  s}$   filters,  and  VVVX
  observations between  years 2016  and 2022  using the  $JHK_{\rm s}$
  filters.  Other colours mark the VVVX areas, observed with $JHK_{\rm
    s}$  filters between  years  2016  to 2022.   Yellow  is the  Disk
  to Longitude $+$230, dark blue is the Low and High Extended Disk, green
  is the Disk to Longitude $+$20, and  in light blue is the Low and High
  Extended Bulge.  A zoomed view of  the image with the  tile names is
  presented in  Appendix A.   Bottom: Schematic representation  of the
  areal  coverage  compared  with  the  other  selected  complementary
  surveys mentioned in Section~\ref{sec:intro}.}
\label{fig:area}
\end{centering}
\end{figure*}

This situation  has improved  in recent  years, with  several projects
studying  the  inner regions  of  the  MW \citep[e.g.][and  references
  therein]{2018ARA&A..56..223B,   2020Msngr.179...31S}.    The   VISTA
Variables     in      the     V\'ia     L\'actea      (VVV)     survey
\citep{2010NewA...15..433M} was  designed to resolve the  3D structure
of the  MW by  searching, precisely  parameterising, and  studying the
distributions of known distance indicators such as RR Lyrae, Cepheids,
and red  clump stars in  the inner  Galaxy.  By using  observations at
near-infrared wavelengths,  VVV observations minimise the  problems of
extinction  and  crowding.  Among  many  results,  the VVV  data  have
enabled the construction of high-resolution extinction and photometric
metallicity                                                       maps
\citep[e.g.][]{2012A&A...543A..13G,2013A&A...552A.110G},          the
discovery of  stellar clusters  \citep[e.g.][]{2014A&A...569A..24B},
and the production of 3D spatial structure maps based on red clump and
RR                             Lyrae                             stars
\citep[e.g.][]{2013ApJ...776L..19D,2013MNRAS.435.1874W}.

In 2016, the  VVV eXtended (VVVX) survey started  operating.  The VVVX
survey was designed to ensure the  long-term legacy of the VVV survey,
characterising the structure  and time domain properties  of the inner
Galaxy.   The   project  is   one  of   seven  large   public  surveys
\citep{2019Msngr.178...10A}  commissioned  by  the  European  Southern
Observatory (ESO).  The VVV+VVVX surveys were awarded about 4200 hours
of observing time  over a timespan of $\sim$\,13 years  at the 4-metre
Visible  and Infrared  Survey  Telescope  for Astronomy  \citep[VISTA;
][]{2006Msngr.126...41E}  telescope at  ESO  Paranal Observatory,  and
were finally completed in March 2023, before the VISTA InfraRed CAMera
\citep[VIRCAM;][]{2006SPIE.6269E..0XD,2010SPIE.7733E...4E} instrument
was decommissioned  from the  VISTA telescope.  Both  surveys combined
cover the  Galactic bulge, as  well as  the adjacent disk  towards the
Galactic quadrants I and IV.

The VVV survey pioneered the  discovery of variable stars, transients,
and a select number of new clusters across a significant region around
the  Galactic centre  and  plane.  VVVX bridges  the  gap between  VVV
findings  and other  surveys, extending  into obscured  regions.  This
aids in  estimating survey  completeness and mapping  distributions of
various tracers from the halo to the Galactic centre.  The VVVX survey
was designed to connect the VVV  survey area with the VISTA Hemisphere
Survey  \citep[VHS;][]{2013Msngr.154...35M}  and the  UKIDSS  Galactic
Plane Survey \citep{2008MNRAS.391..136L}.  VVVX  overlaps with the VST
Photometric H$\alpha$ Survey \citep[VPHAS$+$; ][]{2014MNRAS.440.2036D}
and  the DECam  Plane Survey  \citep[DECaPS; ][]{2018ApJS..234...39S},
providing  complementary  near-IR  imaging   for  those  regions  (see
Fig.~\ref{fig:area}) as well as variability information.  In addition,
by re-observing  the area of the  original VVV, VVVX extends  both the
time baseline  as well as  reaches fainter flux  limits, complementing
other public  optical and far-IR  surveys.  In particular,  the mapped
regions are  located between Galactic longitudes  $l=-130$ degrees and
$l=+20$ degrees, detecting $\gtrsim$\,1.5$  \times 10^9$ point sources
in an  area of  around 1,700  square degrees,  including more  than 50
known globular clusters  and 1,000 open clusters.   The specific goals
of the VVVX survey stated in the original proposal are:

\begin{itemize}

\item To  map the structure  of the optically obscured  populations in
  position and velocity.

\item To find pulsating variable  stars (RR Lyrae, Classical Cepheids,
  Type  2  Cepheids, Miras)  as  distance  indicators probing  the  3D
  structure of the inner MW.

\item  To  physically  characterise  known  and  newly  detected  star
  clusters  (open  star  clusters   as  well  as  globular  clusters),
  measuring  their  distances,  extinctions,  reddenings,  sizes,  and
  estimating their metallicities and ages.

\item To  explore the  stellar populations and  variable stars  of the
  Sagittarius dwarf galaxy located beyond the Galactic bulge.

\item To find rare variable sources  such as transients and WIT (`What
  Is This?')  objects,  and also to identify  the near-IR counterparts
  of  benchmark   high-energy  sources  discovered  by   recent  X-ray
  missions.

\item  To build  a catalogue  with the  classification of  dwarf stars
  beyond the peak of the luminosity function and their companions.

\item To  detect heretoforth unknown  background galaxies and  QSOs in
  the Galactic Zone of Avoidance (ZoA).

\item To  probe the  Galactic structure close  to the  Galactic centre
  using near-IR microlensing.

\end{itemize}

VVVX observations  provide essential  input targets  for spectroscopic
surveys  based on  multiplexing  spectrographs such  as the  SDSS-IV/V
\citep[Sloan            Digital            Sky            Survey-IV/V;
][]{2017AJ....154...28B,2019BAAS...51g.274K},   the   Galactic   4MOST
surveys    \citep[4-metre   Multi-Object    Spectroscopic   Telescope;
][]{2019Msngr.175....3D}\footnote{\tt\tiny{https://www.4most.eu/cms/science/
    galactic-community-surveys/}}    and    MOONS   Galactic    survey
\citep[Multi-Object    Optical    and   Near-infrared    Spectrograph;
][]{2020Msngr.180...18G}.   In  addition,   our  database  complements
measurements from important current and  future space missions such as
the {\it Hubble}  Space Telescope (HST), {\it Gaia},  {\it James Webb}
Space  Telescope (JWST),  {\it Euclid},  and {\it  Nancy Grace  Roman}
Space Telescope.

There is  a variety  of final products,  including deep  $JHK_{\rm s}$
images, multi-band $JHK_{\rm s}$ and multi-epoch $K_{\rm s}$ band time
series  catalogues,  and  proper  motions for  $\gtrsim  1.5$  billion
sources.  Moreover,  VVVX catalogues  millions  of variable  stars,
extend  the  VVV  extinction  and reddening  maps,  and  increase  the
completeness and source density maps, thus presenting a treasure trove
for the whole  astronomical community.  The VVVX  public database will
offer  the  possibility  to  explore  a  wide  variety  of  scientific
objectives, from  those we  listed previously  to new  ones, including
serendipitous discoveries.

\begin{figure*}
\begin{centering}
\includegraphics[scale=0.88]{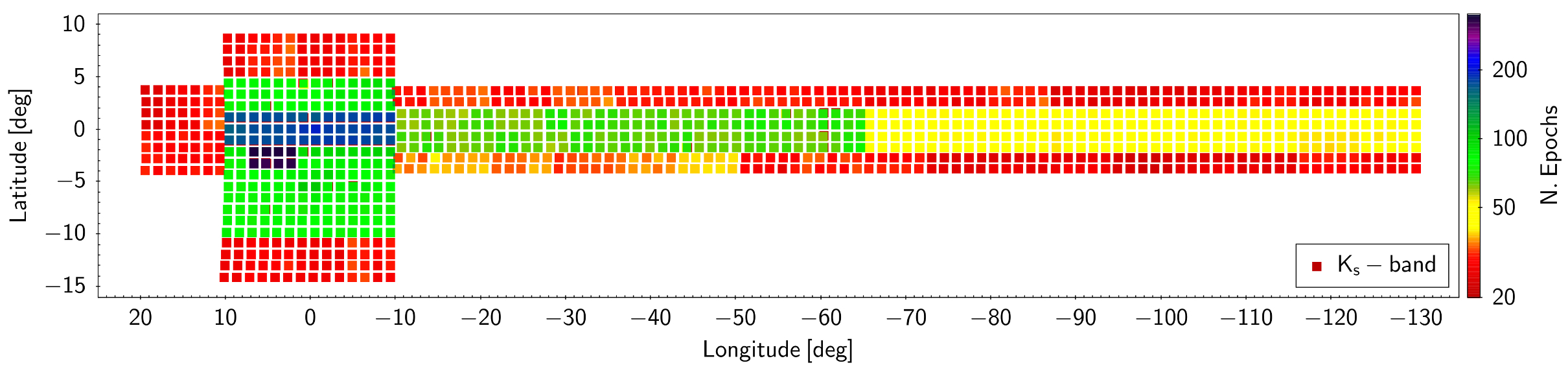}
\caption{Number  of  epochs  in  the  $K_{\rm  s}$  band  observed  by
  VVV+VVVX.   The   mean   baseline   for   the   original   area   is
  $\sim$\,$12$~years, with  up 352  epochs for  selected tiles  in the
  inner bulge. For  the outer bulge and disk the  number of epochs are
  in  the range  $23-106$ for  the VVVX  observations solely,  varying
  according with the observational strategy.}
\label{fig:epochs}
\end{centering}
\end{figure*}

\begin{figure*}
\begin{centering}
\includegraphics[scale=0.55]{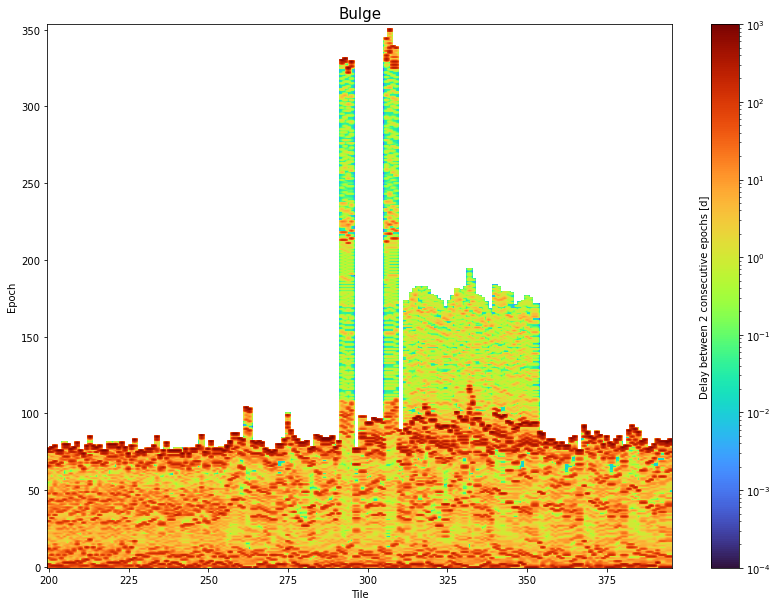}
\includegraphics[scale=0.88]{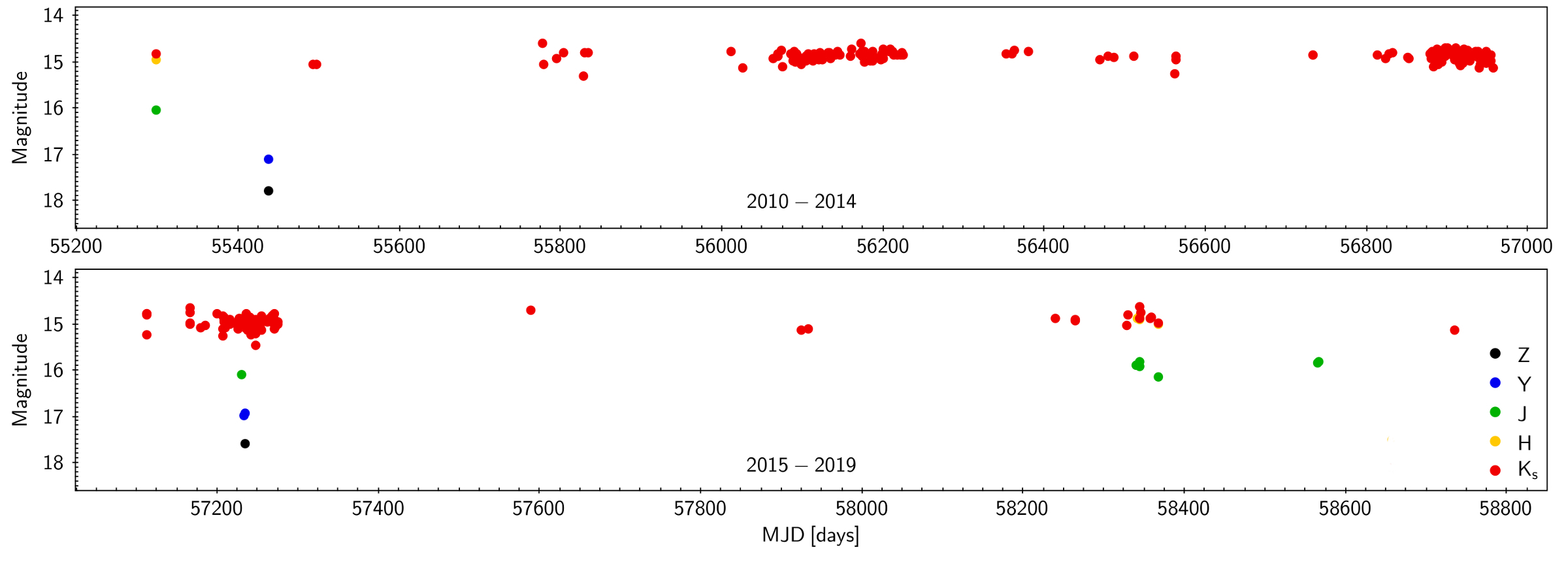}
\caption{Cadence of the VVV+VVVX bulge observations. Top: density plot
  showing the cadence  of the bulge observations in both  VVV and VVVX
  campaigns.  Bottom: light curve example for the source VVV-VIVACE ID
  533558         \citep[$K_{\rm          s}         \sim$\,$14.9$~mag;
  ][]{2022MNRAS.509.2566M},  observed by  VVV+VVVX in  the five  VISTA
  broad band filters in the bulge field b307. There are a total of 363
  observations  in the  five filters  along years  2010 to  2019.  The
  coordinates  for  the target  are  RA,  DEC (J2000)  =  18:00:11.48,
  $-$28:25:13 (corresponding to $l, b=$ 2.0688, $-$2.4904 deg.).}
\label{fig:cadence}
\end{centering}
\end{figure*}

\begin{figure*}
\begin{centering}
\includegraphics[scale=0.85]{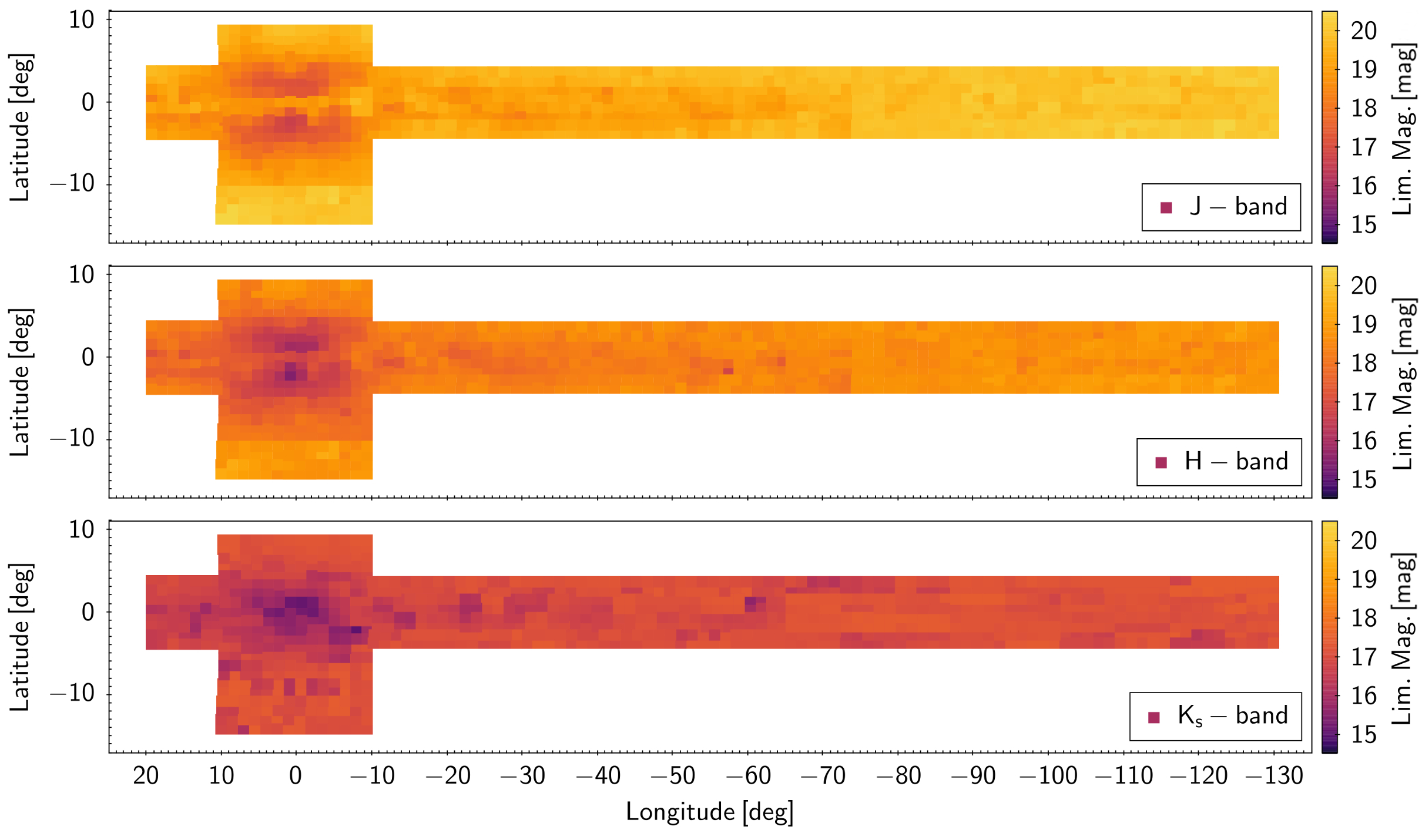}
\caption{5$\sigma$ magnitude limits of the catalogues in the $J$, $H$,
  and $K_{\rm s}$, respectively, from  top to bottom. The colour scale
  is the same in the three panels as shown in the vertical bars at the
  right. For the $J$ and $H$  bands, VVV and VVVX are combined because
  of the lack  of VVVX observation in these bands  in the original VVV
  area.}
\label{fig:maglim}
\end{centering}
\end{figure*}

Due to its larger survey area VVVX provides a more complete picture of
the inner MW  than its predecessor VVV: a deep  bulge map to establish
structure differences  between the  oldest and younger  populations, a
map of the Sagittarius dwarf from its core across the whole bulge, and
a  much  more  extended  disk  map that  probes  star  formation  (SF)
activity, disk  stellar populations,  and spiral arms  structure. VVVX
provides  a  public  multicolour   time  domain  database  within  the
optically hidden MW  regions, including 3D extinction  maps that trace
the  non-stellar  baryonic  matter.   Additionally,  the  VVVX  survey
provides observational constraints for the present-day MW structure as
presently known (e.g., thin and thick disk structure, number of spiral
arms  and  their  locations),  and  even  more  importantly,  provides
insights into the assembly history of the MW.

The  aim of  this  article  is to  describe  the  VVVX survey  design,
observations, data  processing, and final  status of the  VVVX survey,
emphasising the observation strategy  and describing the observed data
available  to the  astronomical  community through  the VISTA  Science
Archive (VSA)  and ESO  Science Archive. We  also describe  some usage
examples of  VVVX data within  the MW and  beyond. In the  ESO Science
Archive the  VVV and VVVX data  are published in the  data collections
identified      by     {\tt{doi.eso.org/10.18727/archive/67}}      and
{\tt{doi.eso.org/10.18727/archive/68}}.

\section{Telescope and instrument}
\label{sec:teles}

The  VVVX observations  were  carried out  with  Visible and  Infrared
Survey Telescope  for Astronomy (VISTA),  an ESO telescope  located at
the Cerro Paranal  Observatory in the Atacama  Desert, northern Chile.
For  all  VISTA  photometric  surveys,  the  VIRCAM  camera  was  used
\citep{2006SPIE.6269E..0XD,2010SPIE.7733E...4E}.  VIRCAM  was equipped
with sixteen  $2048 \times 2048$  science detectors, arranged in  a $4
\times  4$ array,  with a  large  spacing of  90\% and  42.5\% of  the
detector size  along the $X$  and $Y$ axes.  Each  individual detector
covered $\sim$\,$694 \times 694$~arcsec$^2$  on the sky, with $0\farcs
339$ average  pixel scale.   Its filter wheel  was equipped  with five
broad-band  filters ($Z$,  $Y$, $J$,  $H$,  and $K_{\rm  s}$) and  two
narrow-band filters  centred at  0.98 and  1.18~$\mu$m.  For  the VVVX
observations only  $JHK_{\rm s}$ were  used. Table 1 shows  the centre
wavelengths of each filter.

In  all VIRCAM  observations,  each  pointing of  the  telescope is  a
so-called  a  pawprint,  which covers  $0.59$~sq.~deg.   and  provides
partial, coverage  of the  field of view.   By combining  six pawprint
exposures with appropriate offsets, the contiguous coverage of a field
is achieved with at least two  exposures on separate pixels, except at
the edges.  That is called a tile and covers a field of view of $1.109
\times 1.475 = 1.646$~ sq.~deg., the largest unvignetted field of view
in the near-IR regime on 4-m class telescopes.

In 2023, VIRCAM  was decommissioned and will  subsequently be replaced
by the  fibre-fed spectrograph 4MOST  \citep{2019Msngr.175....3D}.  In
fact, various  4MOST surveys will collect  complementary spectroscopic
data to VVV and VVVX.  For details about the telescope and instrument,
we refer the interested  reader to \citet{2015A&A...575A..25S} and the
VIRCAM             instrument            web-pages{\footnote{\tiny{\tt
      http://www.eso.org/sci/facilities/paranal/
      instruments/vircam/}}},   and  the   VISTA/VIRCAM  user   manual
\citep{ivanov+szeifert09}.

\begin{table}
\caption[]{VISTA filters used in the VVVX observations.}
\begin{center}
\label{tab:lambda}
\begin{tabular}{cccc}
\hline \hline 
\noalign{\smallskip}
Filter & ${\rm \lambda}_{\rm eff} (\mu m)$ & $A_X/A_V$  & $A_X/E(B-V)$ \\
\noalign{\smallskip}
\hline
\noalign{\smallskip}
$ J $ & 1.254 & 0.280 & 0.866 \\
$ H $ & 1.646 & 0.184 & 0.567 \\
$ K_{\rm s}$ & 2.149 & 0.118 & 0.364 \\
\noalign{\smallskip}
\hline
\end{tabular}
\tablefoot{The effective wavelengths for  the three VISTA filters used
  in  the  VVVX  observations  are   shown  along  with  the  relative
  extinction for  each filter based on  the \cite{1989ApJ...345..245C}
  extinction law \citep[from][]{2011rrls.conf..145C}.}
\end{center}
\end{table}

\section{Survey area}
\label{sec:area}

The VVV survey  observed $\sim$\,$562$~sq deg in the MW  bulge and the
adjacent southern Galactic plane.  The area was divided in bulge, with
$\sim$\,$300$~sq.~deg.    within   $-10.0^{o}  \lesssim   l   \lesssim
+10.4^{o}$  and $-10.3^{o}  \lesssim b  \lesssim +5.1^{o}$,  hereafter
called  `VVV  Bulge',  and $\sim$\,$220$~sq.~deg.   within  $294.7^{o}
\lesssim  l  \lesssim  350^{o}$  and $-2.25^{o}  \lesssim  b  \lesssim
+2.25^{o}$, hereafter  `VVV Disk' (see Fig.~\ref{fig:area}).   For the
VVV Disk there is  a total of $38 \times 4 = 152$  tiles while the VVV
Bulge is filled up  by $14 \times 14 = 196$ tiles.  The tile sides are
aligned with the Galactic coordinates for coverage optimisation.

The VVVX  survey expanded the  area of  the original VVV  footprint in
both   Galactic   longitude   and    latitude,   with   an   area   of
$\sim$\,$480$~deg$^2$ in the  Galactic bulge plus $\sim$\,1170~deg$^2$
in the inner plane (including the  original VVV), from $l=-130$ deg to
$l=+20$  deg (7\,h\,$<RA<$\,19\,h).   For  contiguous observations  of
large areas,  the covering process  was carried out using  the `Survey
Area         Definition        Tool'         (SADT{\footnote{\tiny{\tt
    https://www.eso.org/sci/observing/phase2/
    SMGuidelines/SADT.html}}}),  which  maximises  the  efficiency  of
VISTA observations by minimising the number of tiles needed to cover a
given area, providing the tile centres and the guide and active optics
stars  necessary for  the  execution of  the  survey Observing  Blocks
(OBs).  The new areas - with their respective tiles - were labelled as
following:

\begin{itemize}
   \item  Low Extended  Disk (Disk-low,  for short):  area with  the
     lowest  latitudes along  the disk,  located within  $+230^{\rm o}
     \lesssim l  \lesssim +350^{\rm o}$  and $-4.5^{\rm o}  \lesssim b
     \lesssim -2.25^{\rm o}$, totalling $\sim$\,266~sq.~deg.  To cover
     the  area, $83  \times 2=166$  tiles were  needed, with  the tile
     names ranging from e0601 to e0766.

    \item High Extended Disk (Disk-high): symmetrical area
      to  the Low  Extended Disk  at higher  latitudes, $+230^{\rm  o}
      \lesssim l \lesssim +350^{\rm o}$  and $+2.25^{\rm o} \lesssim b
      \lesssim  +4.5^{\rm o}$,  comprising $\sim$\,266~sq.~deg.   Tile
      names range from e0767 to e0932.

    \item Disk  to Longitude $+$20 (Disk20): extended the
      disk coverage  to the  north within $+10^{\rm  o}<l<+20^{\rm o}$
      and $|b|\lesssim  4.50^{\rm o}$.   A total of  $7 \times  8 =56$
      tiles were  used to  fill the  area of  $\sim$\,90~sq.~deg.  The
      area  has $\sim$\,90~sq.~deg.   Tile  names in  this region  are
      e0933 to e0988.

    \item  Disk  to Longitude  $+$230 (Disk230): extended
      area   to  the   southern   disk   coverage  within   $+230^{\rm
        o}<l<+295^{\rm o}$  and $|b|\lesssim  2.25^{\rm o}$.   This is
      the  largest  new area,  with  $\sim$\,292~sq.~deg.  A total  of
      $45\times  4=180$ tiles  are in  this region  and labelled  from
      d1001 to d1180.

   \item  Low  Extended  Bulge (Bulge-low): extended  the
     bulge area  within $-15^{\rm  o}<b<-10^{\rm o}$  and $|l|\lesssim
     10^{\rm o}$ with  $\sim$\,90~sq.~deg. in size.  There  is a total
     of $14 \times 4=56$ tiles, labelled from b0401 to b0456.

   \item High Extended  Bulge (Bulge-high):  extended the
     bulge within $+5^{\rm o}<b<+10^{\rm  o}$ and $|l|\lesssim 10^{\rm
       o}$.  The  area has  also $\sim$\,90~sq.~deg.   A total  of $14
     \times 4=56$  tiles filled the area  and are named from  b0457 to
     b0512.

\end{itemize}

The VVVX  footprint along with the  VVV original coverage is  shown in
Fig.~\ref{fig:area}, along with a  comparison with other complementary
surveys,  while  the  tile  names with  the  central  coordinates  are
presented  in Appendix  A.  Both  VVV and  VVVX have  a total  of 1068
tiles.   There  are  no  tiles with  names  d0153-d0200,  b0397-b0400,
b0513-b0600  and   d0989-d1000.   The  absence  of   tile  numbers  is
attributed to the  naming conventions, which are based  on the regions
covered by  the tiles.  The  new VVVX  areas added up  to $\sim$\,2000
hours of observations, split between $\sim$\,450~hrs for $JHK_{\rm s}$
and $\sim$\,1550~hrs for the multi-epoch $K_{\rm s}$.

\section{Observing strategy}
\label{sec:strategy}

The VVVX survey was carried out in service mode, which is the standard
for  all  VISTA observations.   The  observational  blocks (OBs)  were
prepared by the VVVX  team using the P2PP/P2 tools{\footnote{\tiny{\tt
      https://www.eso.org/sci/observing/phase2/p2intro.html}}}     and
sent to ESO for validation and observation at the site.

Observing  blocks for  each tile  were defined  for: a  single $K_{\rm
  s}$ band observation (used in the variability campaign), multicolour
$JHK_{\rm s}$ observations (in a quasi simultaneous schema: $\sim$\,10
minutes for  the sequence $H$ $\rightarrow$  $K_{\rm s}$ $\rightarrow$
$J$),  and additional  $J$ band observations  to be  combined for  the
multicolour   observations.   Due   to   scheduling  constraints   the
multicolour observations had  to be split into  two separate observing
blocks,   including   $JHK_{\rm   s}$   and   $J$ band   observations,
respectively.  We note  that depending on the region, between  2 and 8
adjacent  tiles   were  combined  in  a   concatenation  and  observed
back-to-back.   Minimum  concatenation of  2  tiles  was necessary  to
achieve a  satisfactory background  sky subtraction,  and at  the same
time it reduced observational overheads.

All observations  were carried out using  a 2-point dither due  to the
variable and high  sky background in the near-IR and  to enable cosmic
rays rejection  and pixel  defect correction.   As a  consequence, the
total time per OB for an individual source (e.g., star) and filter is:
DIT  (detector integration  time)  $\times$ NDIT  (number of  detector
integration time)  $\times$ Njitter=2  (number of offsets  executed at
each  of  the  six  pawprint positions)  $\times$  Npaw=6  (number  of
pawprints that together make a tile).  These OB setups lead to a total
time  on the  source in  each  OB of  120s  ($J$), 48s  ($H$) and  16s
($K_{\rm s}$), as presented in Table 2.

The number of epochs observed during  the VVVX campaign in each of the
new areas  ranged between 23 and  50 epochs (see Appendix  A).  In the
original VVV  region, typically 3-12  new epochs were acquired  by the
VVVX campaign,  except in  the inner  Galactic bulge,  where in  a few
tiles up  to 100 additional  epochs have  been secured. In  the region
around the Baade's  window, eight tiles covering  $\sim$\,13 sq.  deg.
in the VVV bulge reached up to 352 epochs, summing up the VVV and VVVX
variability  campaigns.  Fig.   \ref{fig:epochs} shows  the number  of
epochs observed for each tile while Fig. \ref{fig:cadence} illustrates
the cadences for the $K_{\rm  s}$ band variability campaign across the
original VVV bulge area.  Due  to our flexible observing strategy, the
observational sequence for  all tiles is different, and  the ranges of
time separations between  observations are also variable  from tile to
tile.  Therefore, for variability studies, we recommend that each tile
should be treated separately, as, for example, aliasing in the periods
are tile dependent.

$J$ filter only OBs were observed in order to double the exposure time
in that filter.  The deep $J$  imaging corresponds to a time on source
of $\sim$\,4~min,  and due to the  1 hour restriction on  OB duration,
these additional $J$ filter  observations had to be  separate from the
$JHK_{\rm s}$ observations, for a given  tile.  The two $J$ images for
each tile can  be coadded to increase the useful  depth of $K_{\rm s}$
vs  $J-K_{\rm  s}$  colour  magnitude  diagrams.   For  instance,  the
combined $K_{\rm s}$ epochs and the  deep $J$ will reach $J=20.8$ mag,
$K_{\rm s}=20$  mag at  5$\sigma$, which  is three  magnitudes fainter
than the  unreddened bulge main-sequence turnoff.   The densest fields
will  be confusion-limited,  but applying  both point-spread  function
(PSF)  fitting techniques  and  differential photometry  (DIA), it  is
possible to recover  most objects down to $J=19.5$ mag,  $H= 18.5$ mag
and $K_{\rm s}=18$  mag, even in fairly crowded fields.   This is more
than 3 mag fainter than the unreddened RR Lyrae in the Galactic bulge.

Contemporaneous  $JHK_{\rm s}$  epochs give  the colours  for variable
stars (essential  for precise  dereddening of  RR Lyrae  and Cepheids)
while  the  deep  $J$  observations allow  the  construction  of  deep
colour-magnitude  diagrams   reaching  past  the   age-sensitive  main
sequence turn-off.   Because denser fields are  confusion-limited, the
limiting magnitudes vary  along the area, especially  in the innermost
bulge area.  Fig.~\ref{fig:maglim} shows  the limiting magnitudes as a
function of position for all the filters.

\begin{figure}[ht]
\begin{centering}
\includegraphics[scale=0.55]{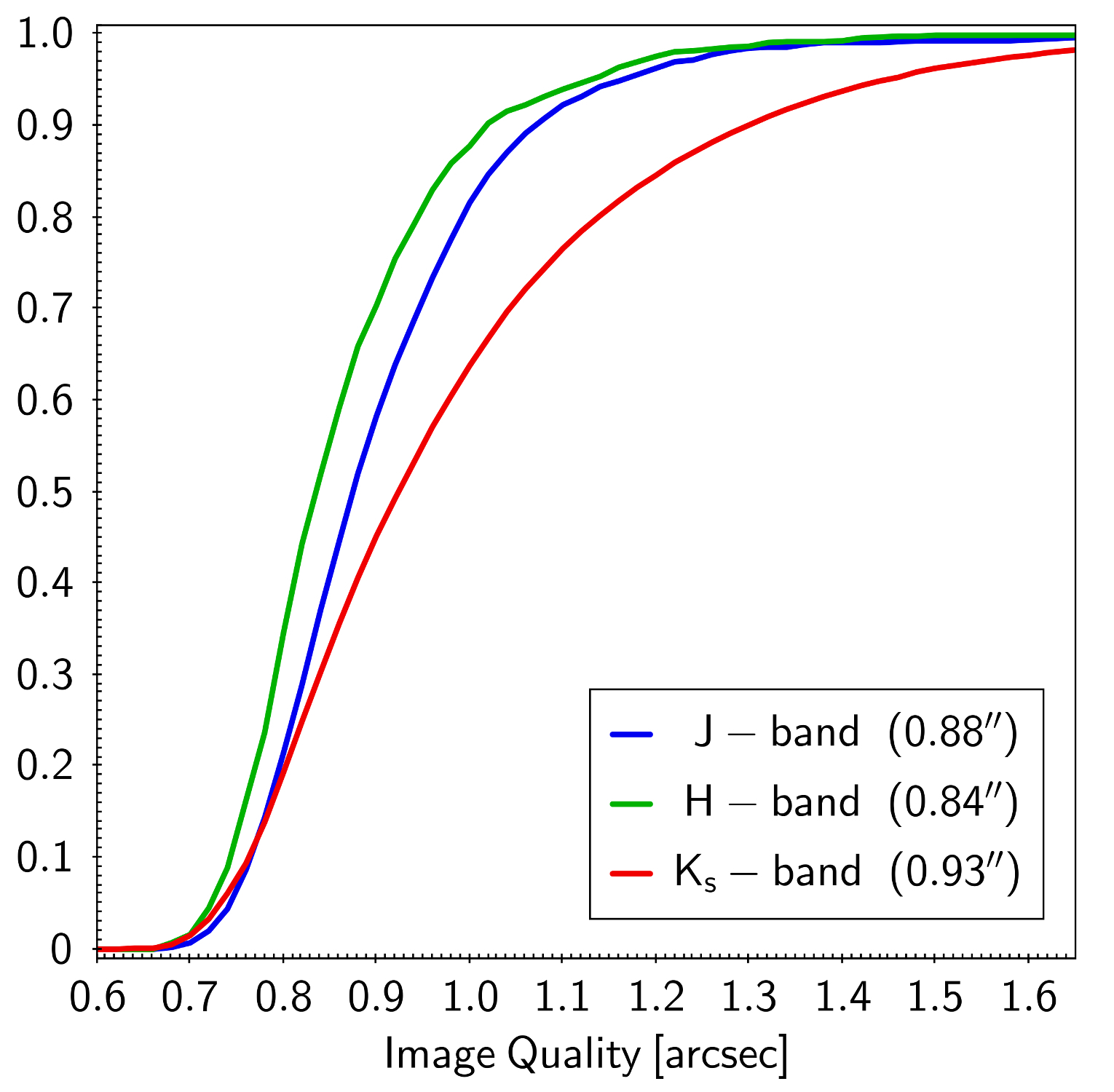}
\includegraphics[scale=0.55]{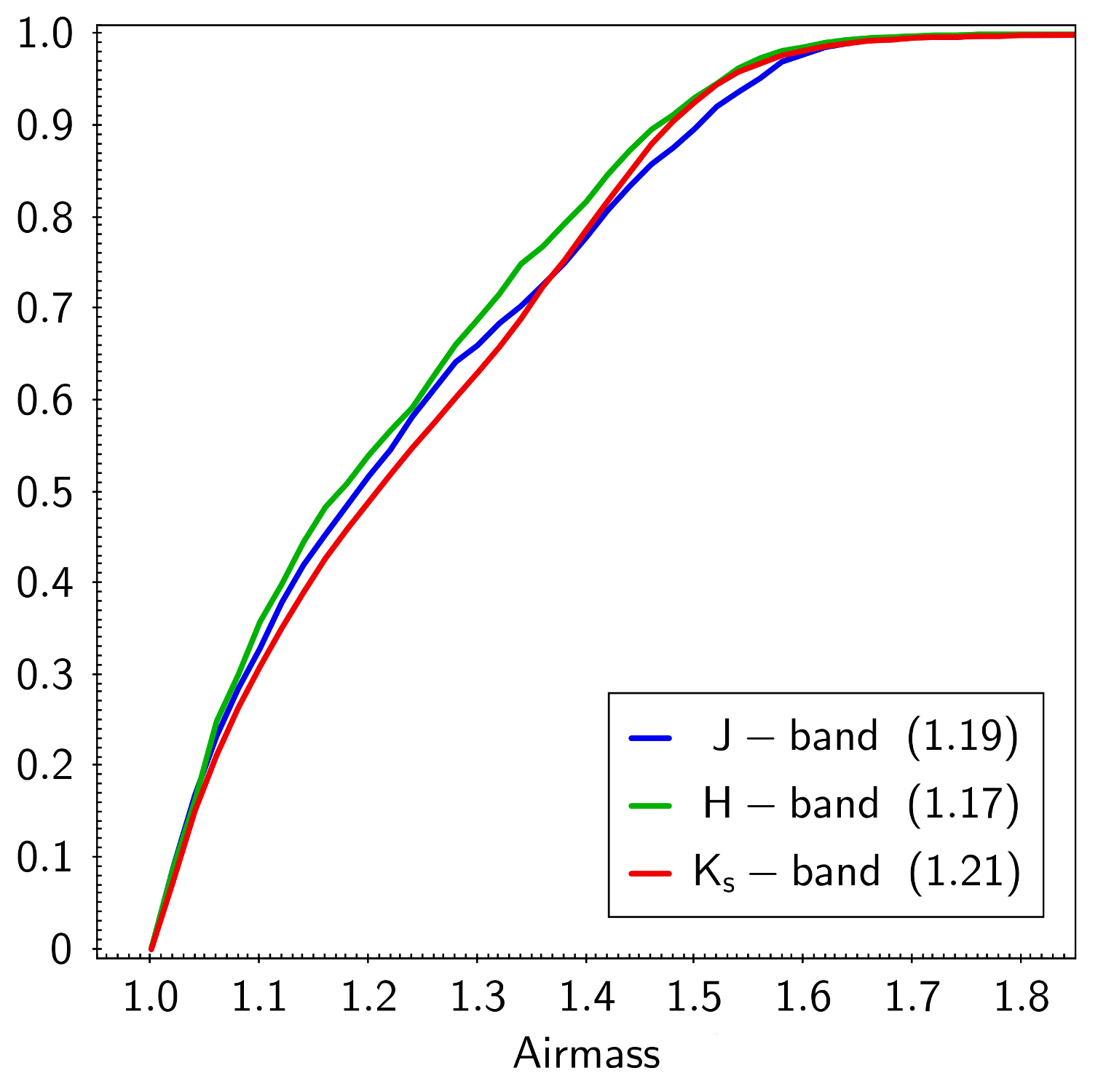}
\caption{Image quality  and airmass  cumulative distributions  for the
  $JHK_{\rm  s}$ VVVX  observations,  in the  top  and bottom  panels,
  respectively. The  median values  of image  quality and  airmass for
  each filter are presented in the legends.}
\label{fig:seeing}
\end{centering}
\end{figure}

\begin{figure}
\begin{centering}
\includegraphics[scale=0.59]{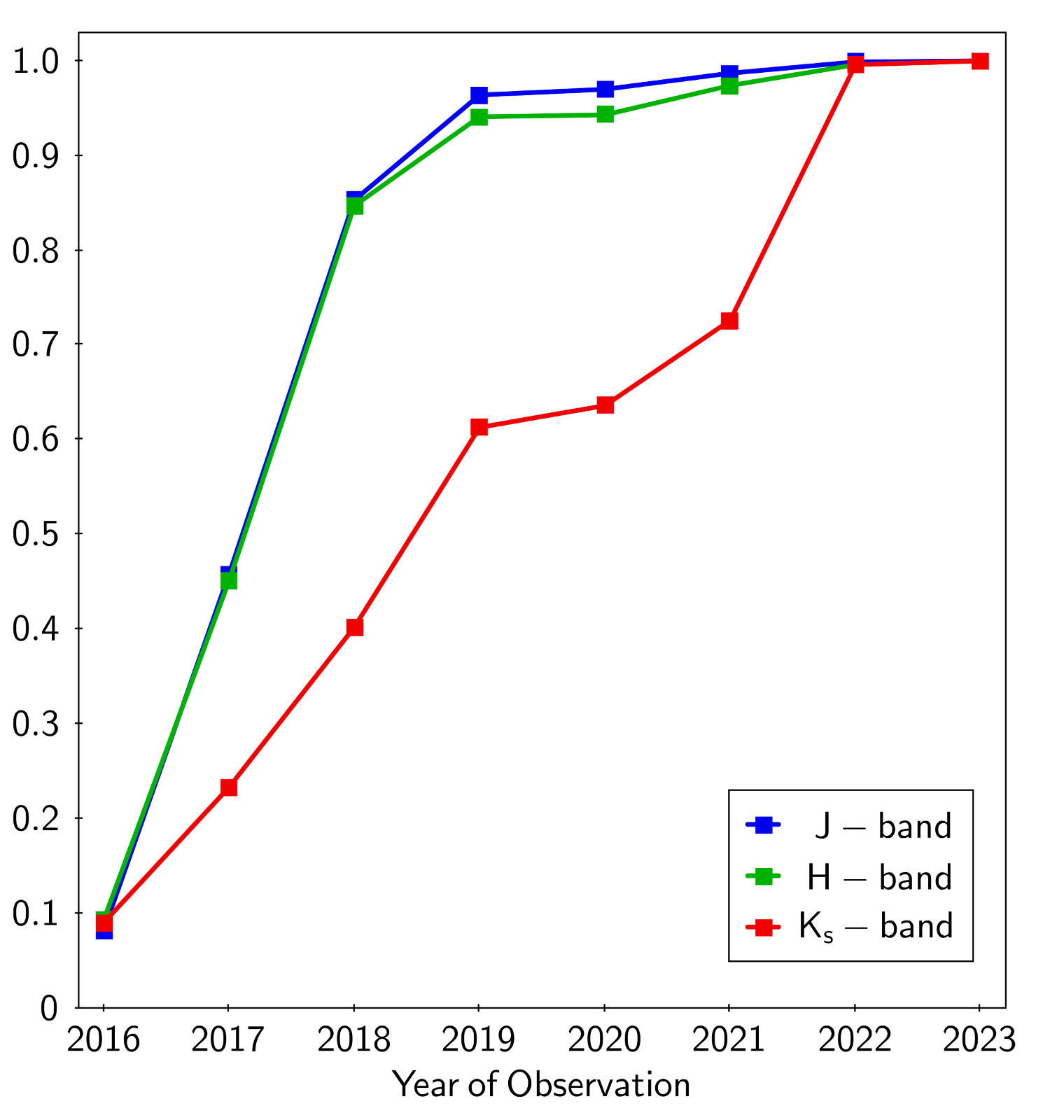}
\caption{Cumulative distributions  for the $JHK_{\rm  s}$ observations
  during the VVVX campaign.}
\label{fig:cumulative}
\end{centering}
\end{figure}

\begin{table}
\begin{center}
  \caption[]{Total exposure time for the  VVVX observations.}
\begin{tabular}{lcccc}
\hline
Filter  &  DIT  & NDIT  & Njitter & Exposure time \\
        &  (s)  &       &         & per pixel (s)    \\
\hline
$J$        &  10  &  3  & 2 & 120 \\
$H$        &  ~~6  &  2  & 2 &  ~~48 \\
$K_{\rm s}$  &  ~~4  &  1   & 2 &  ~~16\\
\hline
\label{table:exp-time}
\end{tabular}
\tablefoot{For each OB the DIT  is the detector integration time, NDIT
  is the number  of the detector integration time, and  Njitter is the
  number   of  offsets   executed  at   each  of   the  six   pawprint
  positions. Njitter=2 applies for most of  the tile area.  Due to the
  tiling process, for a small fraction  of $\sim$8\% of the tiles, the
  Njitter varies  from Njitter=1 (at the  edges) to Njitter=3, 4  or 6
  (in the regions where the pawprints overlap).  For details about the
  tile  areas  covered  for  each  Njitter  value,  see  Fig.   20  in
  VISTA/VIRCAM user manual \citep{ivanov+szeifert09}.}
\end{center}
\end{table}

The $J$  and $H$ observations  were $\gtrsim 85\%$ and  $\gtrsim 95\%$
complete by  the end of  the 2018  and 2019 season,  respectively. The
median image quality in $J$ and $H$ are 0.88'' and 0.84'' for a median
airmass  of 1.19  and 1.17,  respectively, measured  on combined  tile
images (see Fig.~\ref{fig:seeing}). The image quality is measured from
the average  FWHM of sources classified  as bona fide stars  with high
signal-to-noise.   This  value  includes  atmospheric,  telescope  and
instrument related aberrations and is not the same as seeing, which is
an inherent  property of the  atmosphere independent of  the telescope
\citep{2010Msngr.141....5M}.  We  refer to \citet{2015A&A...575A..25S}
for a detailed description of the image quality design and performance
of  the  VIRCAM@VISTA.   In  $K_{\rm s}$,  the  observations  exceeded
$\gtrsim 90\%$ completeness only in 2022, with median image quality of
0.93'' for a  median airmass of 1.21.  In most  cases the observations
satisfied  the atmospheric  turbulence and  photometric image  quality
parameters  and   were  classified   as  completed.    Remaining  $JH$
observations were taken during 2020-2022  and $K_{\rm s}$ in 2020-2023
to complete the planned observations  within the photometric and image
quality parameters.

We summarise  in Table  \ref{table:obs} the  observed areas  along the
years  of the  VVVX  campaign.  Regarding  the  multicolour data,  the
$JHK_{\rm s}$ observations were $\gtrsim  94\%$ complete by the end of
the  2019  season.  The  remaining  observations  were secured  during
$2020-2023$ and comprised repeated observations  of the tiles that did
not pass  the quality control  in the first instance.   The cumulative
distributions    for   the    $JHK_{\rm   s}$    are   presented    in
Fig.~\ref{fig:cumulative}.

The  VVVX  observations  were   pipeline-processed  at  the  Cambridge
Astronomical  Survey Unit  (CASU), using  the VISTA  data flow  system
(VDFS)  pipeline  (Lewis  et  al.   2010).   CASU  also  produces  the
photometric  calibration of  stacked pawprint  images and  tile images
\citep{2018MNRAS.474.5459G}.   Multi-band  catalogues have  also  been
generated  by  the VISTA  Science  Archive  (VSA).  All  tiles,  their
confidence  maps  and  extracted  source lists,  in  addition  to  the
corresponding pawprints, were  processed with version 1.5  of the CASU
pipeline.   In the  catalogues provided  by CASU,  a flag  is used  to
indicate the most probable  morphological classification.  These flags
were  derived  from the  curve-of-growth  analysis  of the  flux  from
different aperture sizes \citep{2004SPIE.5493..411I}. The flags are:\\

  \textbullet~~`$-$1' to denote stellar  objects,

  \textbullet~~`$-$2' to denote borderline stellar,

  \textbullet~~`$-$7' denoting  sources containing  bad pixels,

  \textbullet~~`$-$9' is  used for saturated  stars,

  \textbullet~~`$+$1' is used for non-stellar objects and

  \textbullet~~`0' is used to denote a noise measurement.\\

The  entire reduction  and calibration  process  is the  same as  that
applied  to the  VVV data.  For more  details on  data processing  and
catalogue generation, we  refer the reader to Sections 2  and 3 of the
VVV DR1 paper \citep{2012A&A...537A.107S} and the references therein.

\section{Previous data releases}
\label{sec:dr}

The   first   VVVX    data   release   (VVVX   DR1{\footnote{\tiny{\tt
      https://www.eso.org/rm/api/v1/public/
      releaseDescriptions/130}}}) was  published in February  2019 and
contains data acquired  between July 2016 and August  2017.  The first
release  contains observations  of  about  $590$~ sq.~deg.,  typically
covered in the three filters ($J$, $H$ and $K_{\rm s}$).  DR1 released
a total  of 1660 tile images  that passed quality control,  plus their
associated weight maps and single-band source catalogues. Moreover, it
contains 9960  pawprint images and  their associated weight  maps that
were used  to create the tile  images. The total data  volume is about
1.7      TB      in      compressed     format      (CFITSIO      RICE
compression{\footnote{\tiny{\tt
      http://wsa.roe.ac.uk//qa.html\#compress}}}).

In  June 2021  we publicly  released the  VVVX DR2{\footnote{\tiny{\tt
      https://www.eso.org/rm/api/v1/public/ releaseDescriptions/181}}}
with the data taken between April  2018 and October 2019. They consist
of both  multicolour $JHK_{\rm  s}$ observations and  $K_{\rm s}$ band
variability  data.   The  DR2  has   10776  tile  images,  plus  their
associated  single-band source  catalogues  and weight  maps. For  the
pawprints,  there is  a total  of 64716  images, plus  the single-band
source catalogues  and weight  maps.  The total  data volume  is eight
times larger  than DR1,  reaching 13.4 TB  in compressed  format.  The
uncompressed data  volume of the  VVVX DR1  plus VVVX DR2  is $\gtrsim
27$~TB.

The $K_{\rm s}$  band taken prior to April 2018  was not released with
the VVVX DR1, which contained data only from the multicolour $JHK_{\rm
  s}$  OBs and  $J$-only OBs.   $JHK_{\rm s}$  and $J$-only  OBs taken
during 2020-2022, as well as $K_{\rm  s}$ from 2020 to 2023, represent
a smaller  part of  the total  volume. These  data had  been gradually
processed by  CASU and VSA, and  ingested to the ESO  Archive, with no
announcement of a public data  release.  These remaining data complete
the  release to  the astronomical  community of  all the  observations
planned by VVVX within the established quality criteria. In total, the
VVV+VVVX  survey   observations  comprise  $>350,000$   images,  which
including images, catalogs  and confidence maps, they make  a total of
more than $10^6$~files.

The VVVX data, as described above,  can be downloaded directly via the
ESO    Archive\footnote{\tiny{\tt   http://archive.eso.org/cms.html}}.
Various options are  available for how the data can  be obtained, from
simple cone searches to more complex options. In the ESO Archive it is
also possible to browse the VVVX  data using an Aladin Sky Atlas tool,
and by clicking  and selecting, determine which data  among images and
catalogues are of interest for direct download.

\section{Advanced data products}
\label{sec:advanced}

The  images and  catalogues described  in the  section above,  already
available  via the  VSA and  ESO  Archive, are  being complemented  by
advanced products,  currently under  preparation by VVVX  science team
members.   As  for  VVV,  a  key product  is  the  extraction  of  PSF
photometry, which is more robust and accurate than aperture photometry
data  for the  regions  of high  crowding  of stars,  such  as in  the
innermost  MW bulge  and plane,  and for  the study  of star  clusters
\citep[e.g.][]{2018A&A...619A...4A}.

PSF-based catalogues are compiled for each VVVX tile, however measured
from  the  stacked  pawprint  images,  where  codes  such  as  DAOPHOT
\citep{1987PASP...99..191S},  DOPHOT \citep{1993PASP..105.1342S},  and
SExtractor \citep{1996A&AS..117..393B} work with better precision.

As  well  as  for  the   VVV  Survey,  various  authors  will  provide
independently produced PSF based  source catalogues of the multicolour
data  \citep[e.g.][]{2018A&A...619A...4A,2019A&A...629A...1S},  and
the   variability   campaign   \citep[e.g.][]{2017A&A...608A.140C,
  2018MNRAS.474.1826S}.  Specific details will be published elsewhere.
The catalogues  will be  available through  the VSA\footnote{\tiny{\tt
    http://vsa.roe.ac.uk/}}, using the same schema as for aperture and
image data.

The $K_{\rm  s}$ band variability data  allows for a variety  of uses,
from constructing light curves to  proper motion measurements. For the
VVV  area,  our  team  published a  number  of  catalogues  containing
hundreds of  thousands candidate of  variable stars, for  example: the
VVV       Near-IR      Variability       Catalogue      \citep[VIVA-I;
][]{2020MNRAS.496.1730F},  the  Near-IR  Catalogue of  known  variable
stars    \citep{2021A&A...647A.169H},   and    the   VIRAC    Variable
Classification Ensemble \citep[VIVACE; ][]{2022MNRAS.509.2566M}.

The  extinction  in $K_{\rm  s}$  filter  of  VIRCAM  is an  order  of
magnitude lower than  in optical.  Hence, the VVVX  proper motion data
are complementary to those of  {\it Gaia} in highly extincted Galactic
disk and  bulge regions, where the  optical {\it Gaia} data  result in
significantly  reduced  depth.  Over  the  area  of the  original  VVV
region, the $K_{\rm s}$ band data can be combined to increase the time
baseline to more than a decade  ($\gtrsim 2010-2020$, or even 2022 for
selected bulge  fields), thus  increasing the  accuracy of  the proper
motion  measurements.  For  the new  extended area,  the time  base is
approximately $4-5$ years.

Proper-motion catalogues will be  incorporated into the VISTA Infrared
Catalogue  2  (VIRAC2, in  preparation),  the  latest version  of  the
original       VISTA       Infrared      Catalogue       \citep[VIRAC;
][]{2018MNRAS.474.1826S}.   \citet{2023A&A...677A.185L}  compared  the
VIRAC2, {\it Gaia} DR3 and HST  proper motions in a few fields towards
the Galactic bulge, with different  stellar crowding levels.  The test
showed  that VIRAC2  proper motions  have more  reliable uncertainties
than {\it Gaia} DR3 and are comparable to HST in dense fields, such as
globular clusters and the Galactic bulge. The shorter time base of the
new VVVX areas should influence the  data quality, so we should expect
larger fractional  uncertainties for more distant  and slower sources.
VIRAC2 has been uploaded to ESO archive where it will be available for
Virtual Observatory (VO) TAP queries.

Aside from the aperture and PSF  photometry for the point sources, the
images  also  contain  extended  sources.  The  optical  detection  of
extragalactic sources beyond the MW is hampered in the ZoA, where the
stellar  crowding  and Galactic  absorption  are  rather extreme  (see
Subsection \ref{sub:background}).

\begin{table*}
\begin{center}
\caption[]{Summary of the VVVX observational campaign.}
\begin{tabular}{lccccccccccc}
\hline Area Name  & Area & No.  of & RA Range  & 2016 & 2017  & 2018 &
2019 & 2020 & 2021 & 2022 & 2023 \\
\vspace{3 pt}
                      &  (deg$^2$) & Tiles   & (hours)   &      &      &      &      &      &      &      &       \\
\hline
VVV bulge          &  313      & 196      & 17h$-$19h &  ~~~~~$K_{\rm  s}$ & ~~~~~$K_{\rm  s}$ &  $JHK_{\rm  s}$   & ~~~$JK_{\rm  s}$  & ~~~~-- & ~~~~~$K_{\rm  s}$ &  ~~~~~$K_{\rm  s}$  &  ~~~~-- \\   
VVV disk           &  232      & 152      & 12h$-$19h &  ~~~~~$K_{\rm  s}$ & ~~~~~$K_{\rm  s}$ & ~~~~~$K_{\rm  s}$ & ~~~~~$K_{\rm  s}$ & ~~~~-- & ~~~~--            &    ~~~~~$K_{\rm  s}$ & ~~~~-- \\
Disk Long. $+$20   &   90      &  56      & 18h$-$19h &  ~~~~~$K_{\rm  s}$ &  $JHK_{\rm  s}$   & ~~~~~$K_{\rm  s}$ & ~~~~~$K_{\rm  s}$ & ~~~~-- & ~~~~--            &    ~~~~~$K_{\rm  s}$ & ~~~~-- \\
Disk Long. $+$230  &  292      & 180      & 07h$-$12h &  --            &  --           &  $JHK_{\rm  s}$   &  ~~~~~$K_{\rm  s}$ &  ~~~~~$K_{\rm  s}$ &  ~~~~~$K_{\rm  s}$ & ~~~~~$K_{\rm  s}$ &    ~~~~~$K_{\rm  s}$  \\
Low Ext. Bulge     &   90      &  56      & 18h$-$19h & $JHK_{\rm  s}$    &  --            &    ~~~~~$K_{\rm  s}$  &    ~~~~~$K_{\rm  s}$ & ~~~~-- &~~~~ --     & ~~~~--        & ~~~~--      \\
High Ext. Bulge    &   90      &  56      & 17h$-$18h &    ~~~~~$K_{\rm  s}$ &  $JHK_{\rm  s}$  &    ~~~~~$K_{\rm  s}$  &    ~~~~~$K_{\rm  s}$ & ~~~~-- & ~~~~-- &    ~~~~~$K_{\rm  s}$ & ~~~~--     \\
Low Ext. Disk      &  266      & 166      & 07h$-$18h &    ~~~~~$K_{\rm  s}$ &  $JHK_{\rm  s}$  &  $JHK_{\rm  s}$   & $ JHK_{\rm  s}$   &    ~~~~~$K_{\rm  s}$ &    ~~~~~$K_{\rm  s}$ &    ~~~~~$K_{\rm  s}$  &    ~~~~~$K_{\rm  s}$  \\
High Ext. Disk     &  266      & 166      & 07h$-$17h &    ~~~~~$K_{\rm  s}$ &  $JHK_{\rm  s}$  &  $JHK_{\rm  s}$   & $ JHK_{\rm  s}$   &    ~~~~~$K_{\rm  s}$ &    ~~~~~$K_{\rm  s}$ &    ~~~~~$K_{\rm  s}$  &    ~~~~~$K_{\rm  s}$  \\
\hline
\label{table:obs}
\end{tabular}
\tablefoot{$J$ and  $H$ observations were  97\% and 94\%  completed by
  the end of  2019 season. The remaining observations  were taken from
  2020 to 2023, mostly repeated observations. The $K_{\rm s}$ band was
  $\gtrsim  99\%$  completed by  the  end  of  the 2022  season.  Only
  residual $K_{\rm s}$ observations,  comprising $\lesssim 1\%$ of the
  total,   were  taken   in  year   2023  to   complete  the   planned
  observations.}
\end{center}
\end{table*}

\section{Examples of data usage: Tile e1084, Carina Nebula, and NGC 3324}
\label{sec:e1084}

\subsection{Multicolour images and catalogues}

One  of  the first  images  released  by the  JWST  is  of the  Carina
Nebula{\footnote{\tiny{\tt{https://www.nasa.gov/image-feature/goddard/2022/nasa-s-
      webb-reveals-cosmic-cliffs-glittering-landscape-of-star-birth/}}}. The
  nebula is located within the Disk  to Longitude $+$230 area, in tile
  e1084, which has central coordinates  RA, DEC (J2000) = 10:30:32.86,
  $-$58:39:41.8,  corresponding  to   $l,  b=-$74.524725,  $-$0.649149
  deg. The $JHK_{\rm  s}$ observations of tile e1084  were carried out
  on  March   19,  2018.    Table  \ref{table:e1084}   summarises  the
  observational  log.  A  composite false  colour image  combining the
  $JHK_{\rm  s}$   images  can  be   produced  with  the   Aladin  Sky
  Atlas\footnote{\tiny{\tt https://aladin.cds.unistra.fr/}}  for e1084
  using the  WCS data in the  reader of each image  to determine their
  location, rotation and image scale.

\begin{table*}
\begin{center}
\caption[]{Multicolour observations of tile e1084.}
\begin{tabular}{lcccccccc}
\hline
  Band      &   Obs. date          & Airmass &  Exp. time &  Seeing &   ZP   &  Maglim &  N-sources &   N-flags \\
\vspace{3 pt}
            &    \small{(2018-03-20UT)}   &           &  (s)     &  ($''$) &  (mag) & (mag)   &          &  ~~$-1$~~~~~~~~~~~$+1$~~~~~~~~~~$-2$~~~~~~~~(sum)  \\
\hline
  $H$       &  06:35:58.7520      & 1.494   &  24.00   &  0.72   &  23.71 &  18.72  & ~~\,998,946 &   605,248~~314,151~~65,572~~  (98.6\%)   \\
 $K_{\rm s}$ &  06:41:43.4729      & 1.514   &   8.00   &  0.73   &  22.88  &  17.17  & ~~\,592,401 &   338,805~~229,305~~18,006~~  (98.9\%)   \\ 
  $J$       &  06:45:19.4624      & 1.527   &  60.00   &  0.74   &  23.61  &  19.95  & 1,013,278   &  641,619~~288,262~~72,230~~  (98.9\%)    \\
\hline
\label{table:e1084}
\end{tabular}
\tablefoot{Observational  log   and  catalogue  information   for  the
  multicolour observations  of the VVVX tile  e1084.  All observations
  were labelled as `Completed' and  `ESO Grade A'.  The catalogues
  are dominated  by stellar ($-1$), non-stellar  ($+1$) and borderline
  stellar  ($-2$) sources,  accounting for  $\sim$\,99\% of  the total
  number in each catalogue.}
\end{center}
\end{table*}

In Fig.   \ref{fig:carina} we compare  the three colour  $JHK_{\rm s}$
image  obtained with  VISTA/VIRCAM with  its JWST/NIRcam  counterpart.
Both images  have $\sim$\,$9 \times  7$ arcmin size, with  a different
orientation than  equatorial or Galactic coordinates.   JWST composite
image  comprises separate  exposures containing  F090W, F187N,  F200W,
F335M, F444W, F470N NIRCam narrow  and wide filters. These filters are
part  of  both  the  short  and  long  wavelength  channels,  covering
wavelengths from 0.901 to 4.707 microns, corresponding to pixel scales
of  $0.031''$/pixel  and  $0.063''$/pixel   for  the  short  and  long
wavelength channels, respectively. Although the JWST images were taken
at a broader range of  longer wavelengths ($0.90 - 4.70~\mu$m compared
to $1.25  - 2.15~\mu$m for VISTA),  the VVVX images are  comparable in
quality  for  relatively  bright  point  sources,  while  the  nebular
structure  definition is  clearly  superior in  the  JWST.  For  faint
sources, the  higher resolution of  JWST compared to VISTA  allows the
first to reach several magnitudes deeper at the same wavelengths.

Near the  Carina Nebula (Fig.   \ref{fig:carina}) we find NGC  3324, a
stellar cluster first  described by \citet{2005A&A...438.1163K}, based
on PPMLX and 2MASS data. As  described before VVVX offers NIR data not
only  with  a  much  better   resolution,  but  also  reaches  several
magnitudes deeper. In  Fig. \ref{fig:cluster} we present  the $J$, $H$
and $K_{\rm  s}$ band images for the  cluster. It is possible  to note
the  difference in  the gas  transparency towards  longer wavelengths,
especially in the lower part of the images where the gas concentration
is higher,  increasing the  number of background  sources that  can be
observed. 

We created  a multicolour catalogue  for tile e1084  by cross-matching
the  $J$,  $H$  and  $K_{\rm  s}$ band  catalogues  provided  by  CASU
(aperture   photometry,  see   Table   \ref{table:e1084}).   For   the
cross-match  procedure  we   made  use  of  STILTS{\footnote{\tiny{\tt
    https://www.star.bris.ac.uk/$\sim$mbt/stilts/}}}~\citep{2006ASPC..351..666T},
allowing  a tolerance  of  $1"$  between the  sky  coordinates of  the
detected  sources, resulting  in a  $JHK_{\rm s}$ band  catalogue with
568k sources,  which is limited  by the  shorter exposure time  in the
$K_{\rm  s}$ band   compared  with   the  other  filters   (see  Table
\ref{table:exp-time}).   Multiepoch  $K_{\rm  s}$ band images  can  be
coadded  to provide  a deeper  catalogue; however,  that has  not been
applied  here.  In  using  the photometric  flags  and selecting  only
stellar sources  in all 3  bands (flag `-1', see  previous section),
the number of sources in the e1084 catalogue dropped from 568k to 262k
sources.  The $J \times (J-K_{\rm s})$  CMD of stellar sources in tile
e1084 is  presented in  Fig.  \ref{fig:cmd}.   We have  also similarly
prepared CMDs  for the  region of  Carina and NGC  3324, for  the same
areas presented in Figs.  \ref{fig:carina} and \ref{fig:cluster}.

In order to demonstrate the importance  of the PSF photometry, we also
present CMDs using PSF data for the same regions mentioned above: tile
e1084,  the Carina  Nebula region,  and  NGC 3324.   We are  currently
finishing the  PSF atlas for  the whole VVVX area  (Alonso-Garc\'ia et
al. in prep.), building on our previous experience extracting the PSF
photometry       from      the       original      VVV       footprint
\citep{2018A&A...619A...4A}.   A   preliminary  version  of   the  PSF
catalogues was  used to build the  CMDs of these areas  of interest in
Fig.  \ref{fig:cmd}.

Although there  is excellent  agreement between  the aperture  and PSF
photometries,  a larger  number of  sources in  the PSF  catalogues is
evident since our algorithm to extract  the PSF photometry uses a less
conservative limit than the  CASU aperture photometry. This difference
becomes larger  for the  more crowded fields  over the  Galactic plane
where the PSF  is far more efficient. The VVVX  PSF catalogues for the
entire area will be described in  Alonso-Garcia et al. (in prep.)  and
publicly released to the community through VSA.

\begin{figure*}
\sidecaption
  \includegraphics[width=12cm]{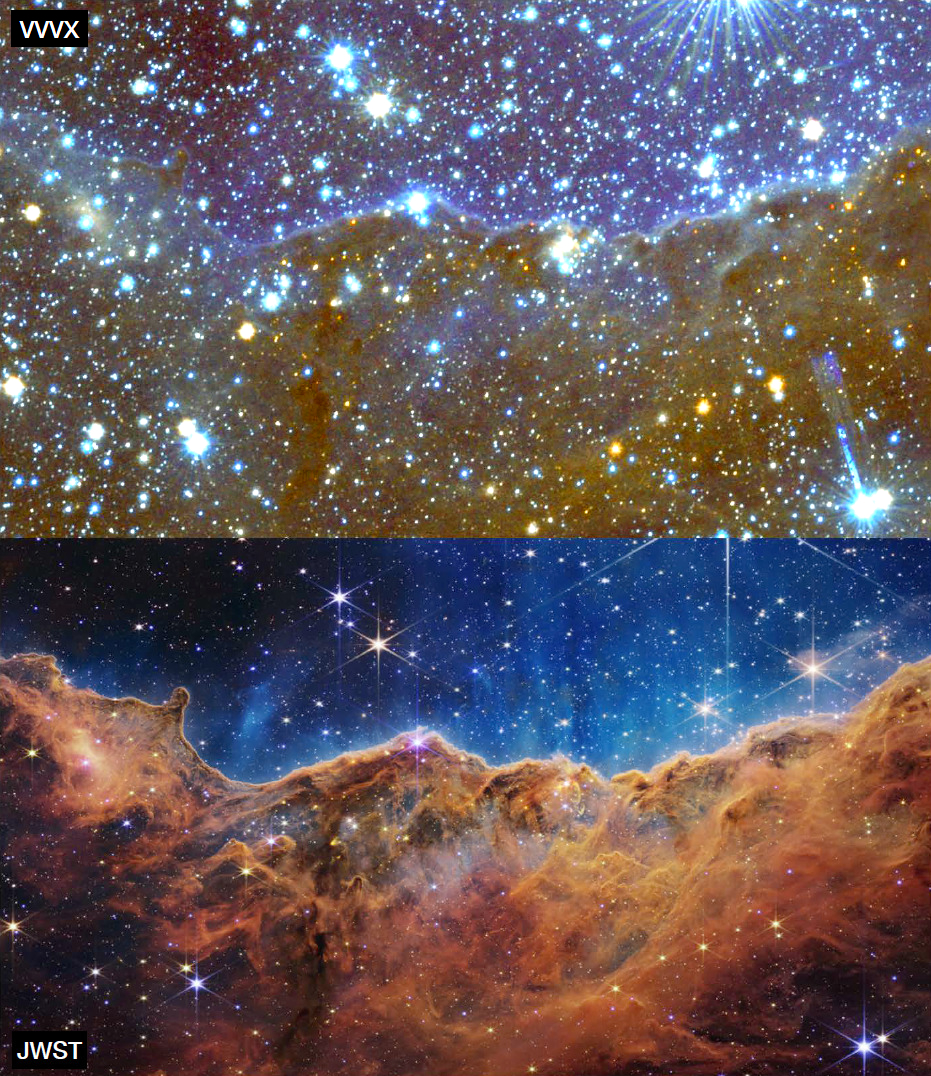}
     \caption{VVVX composite image of  the Carina Nebula compared with
       the JWST  image of the same  field. The VVVX image  (top) is in
       false colour and based on  the $JHK_{\rm s}$ observations while
       the JWST image (bottom) comprises separate exposures containing
       F090W, F187N, F200W, F335M, F444W and F470N NIRCam filters. The
       Carina  Nebula is  located towards  tile e1084  in the  Disk to
       Longitude  $+$230  region  of  VVVX,  with  the  $JHK_{\rm  s}$
       observations  secured  on  March  19, 2018.   The  images  have
       $\sim$\,$9  \times 7$  arcmin size  and are  oriented with  the
       north to the right and east  to the top.  Credits (JWST image):
       NASA, ESA, CSA, and STScI, J. DePasquale (STScI).}
     \label{fig:carina}
\end{figure*}

\begin{figure*}
\centering
\includegraphics[scale=0.18]{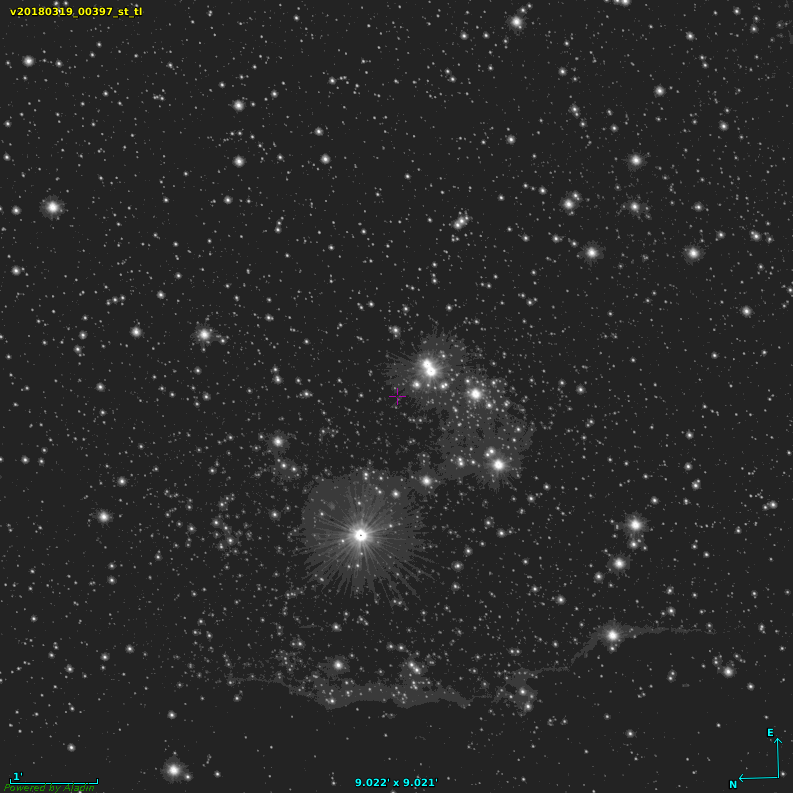}
\includegraphics[scale=0.18]{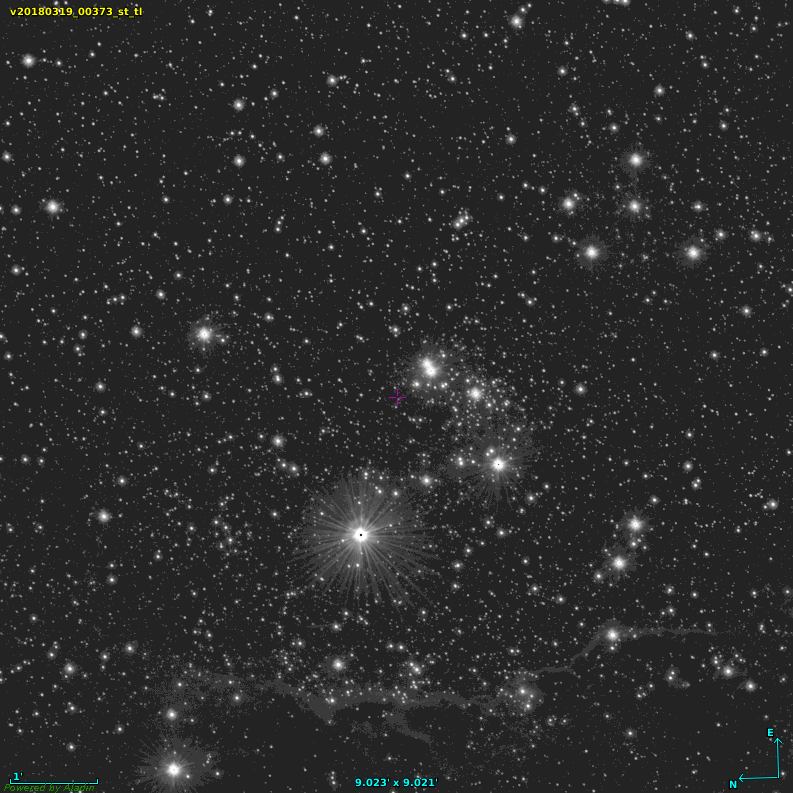}
\includegraphics[scale=0.18]{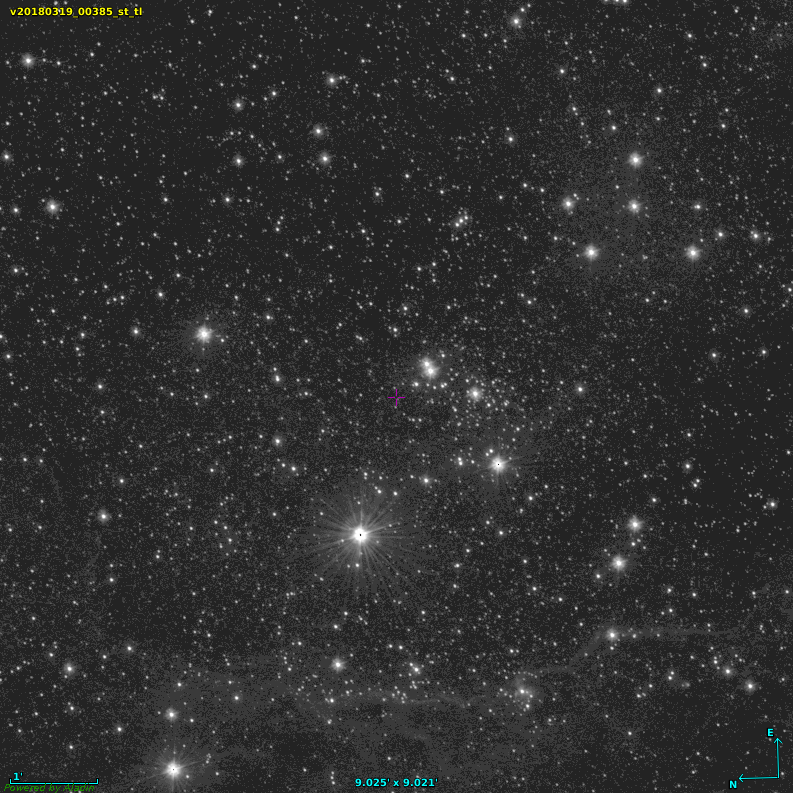}
\caption{Individual  $J$ (left),  $H$  (centre)  and $K_{\rm  s}$ band
  images (right) of the star cluster NGC 3324.  The image is centrered
  at RA/DEC (J2000)=10:37:21.8, $-$58:36:54,  with $9$ arcmin side and
  oriented in Galactic coordinates.}
\label{fig:cluster}
\end{figure*}

\begin{figure*}
\centering
\includegraphics[scale=0.8]{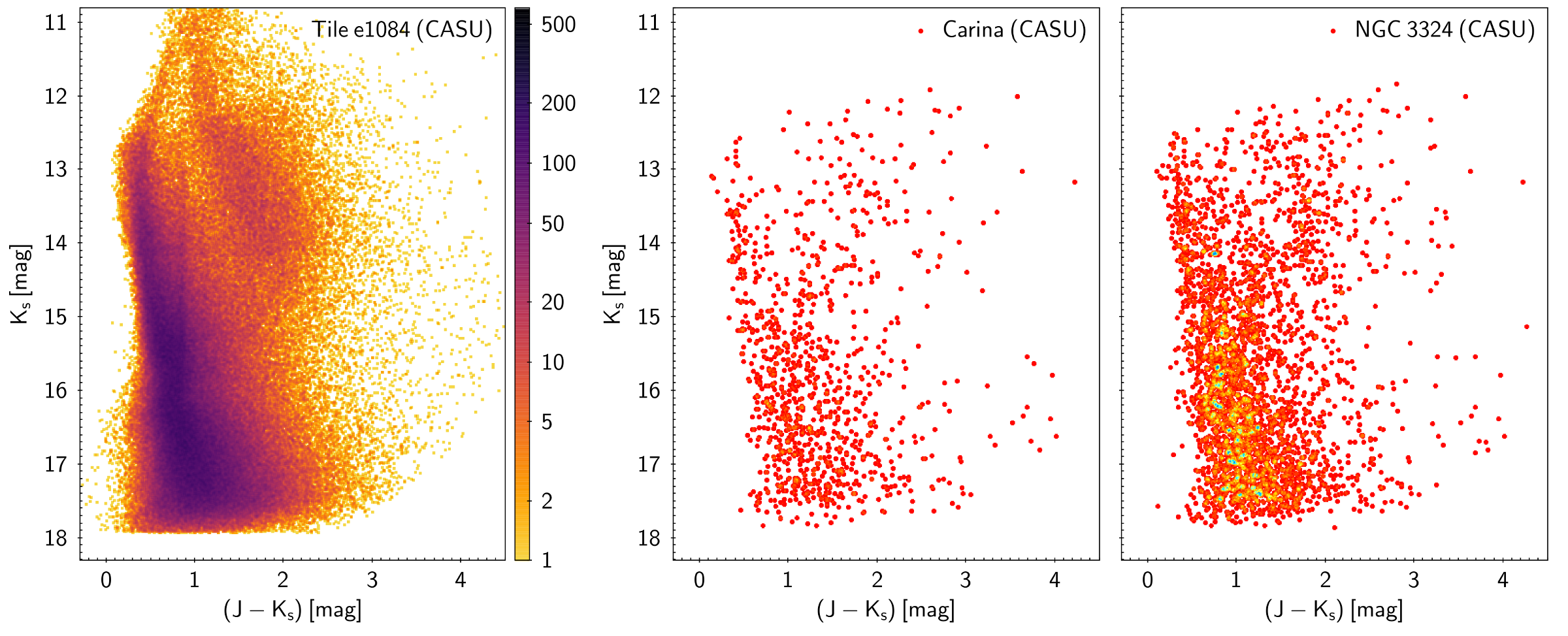}
\includegraphics[scale=0.8]{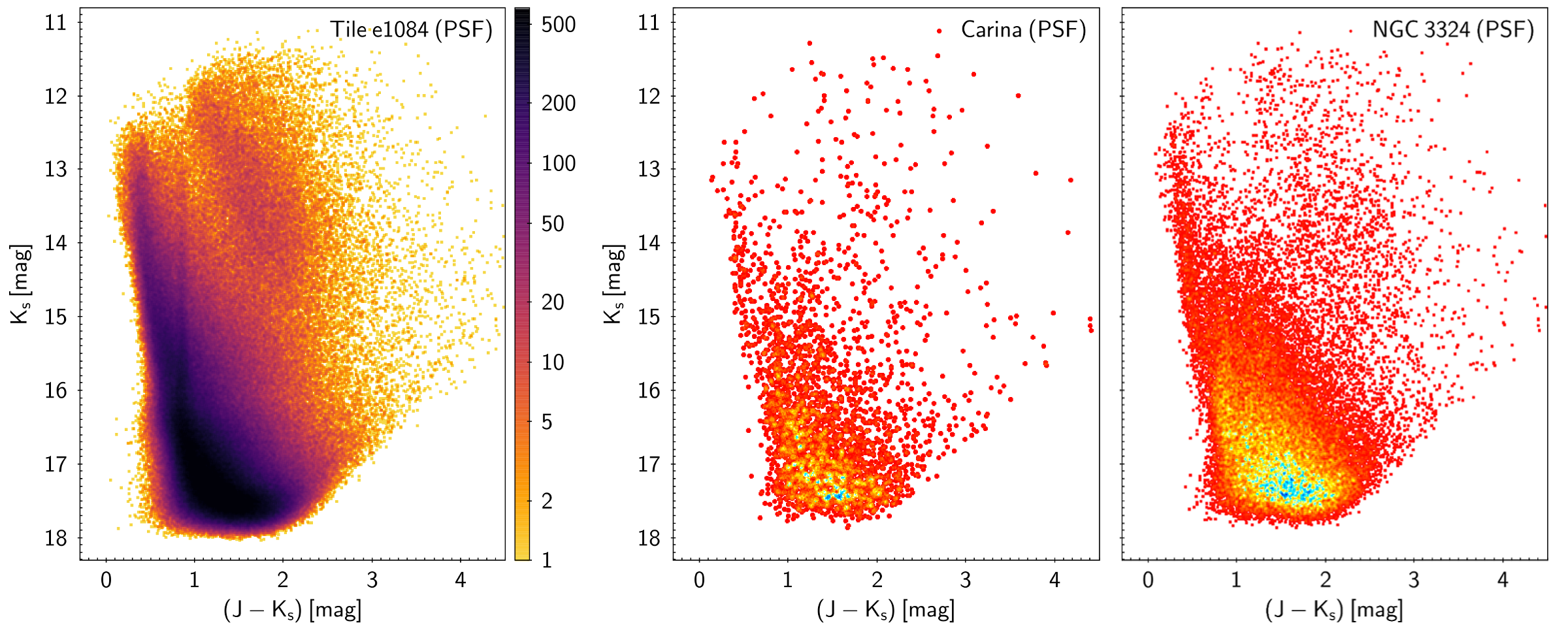}
\caption{Comparative $K_{\rm s} \times (J-K_{\rm s})$ colour-magnitude
  diagrams. Top:  CASU aperture  photometry for stellar  sources (flag
  $-$1) for the entire e1084 tile (left), for the region of the Carina
  Nebula shown in Fig. 7 (centre) and for the stellar cluster NGC 3324
  (right  panel, 6  arcmin  radius  area). The  sharp  end at  $K_{\rm
    s}=17.9$ is caused by the stellar flag criterion applied.  Bottom:
  the CMD for the  same areas in the top using  PSF photometry. In the
  CMDs for the entire tile, the  narrow and vertical structure seen at
  $(J-K_{\rm  s}) \sim  0.9$ is  the sequence  of unreddened  nearby M
  dwarfs \citep[e.g.][]{2018A&A...619A...4A, 2022A&A...660A.131M}.}
\label{fig:cmd}
\end{figure*}

\begin{figure*}
\centering
\includegraphics[scale=0.9]{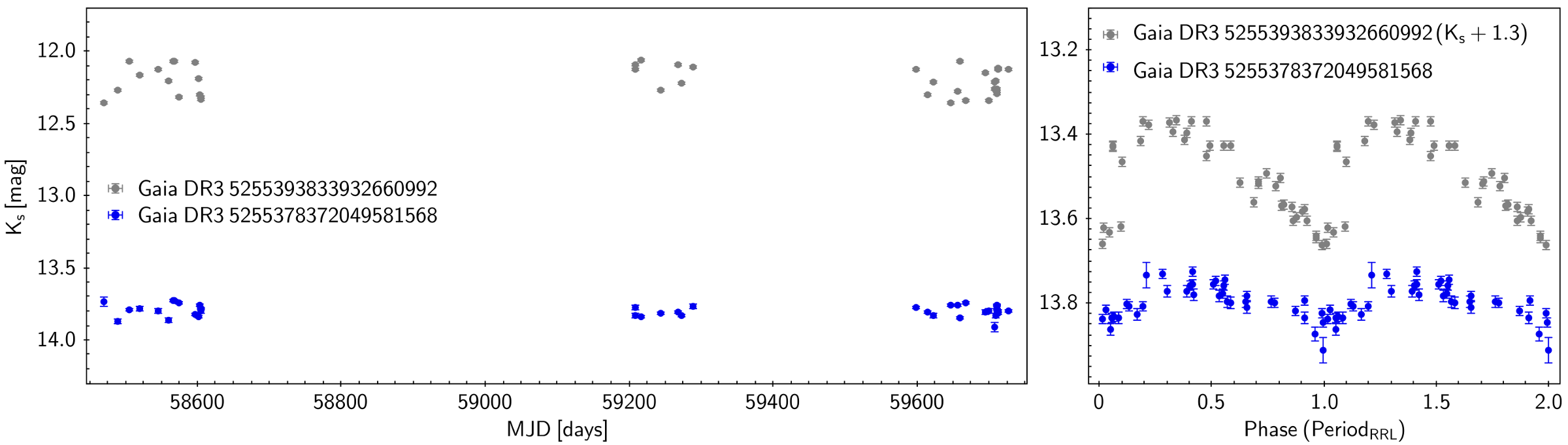}
\caption{VVVX  $K_{\rm s}$ band  light curves  for two  {\it Gaia}  RR
  Lyrae found in  tile e1084. {\it Gaia} DR3  5255393833932660992 is a
  fundamental   mode    RR   Lyrae    (RRab)   with   a    period   of
  $P\sim$\,0.476~days while  {\it Gaia}  DR3 5255378372049581568  is a
  first-overtone  RR Lyrae  (RRc)  with  $P\sim$\,0.238~days.  In  the
  right  panel  the  phase  folded  light  curve  of  {\it  Gaia}  DR3
  5255393833932660992  is  arbitrarily  shifted   by  $+$1.3  mag  for
  visualisation purposes.}
\label{fig:rrl}
\end{figure*}

\begin{figure}
\centering
\includegraphics[scale=0.8]{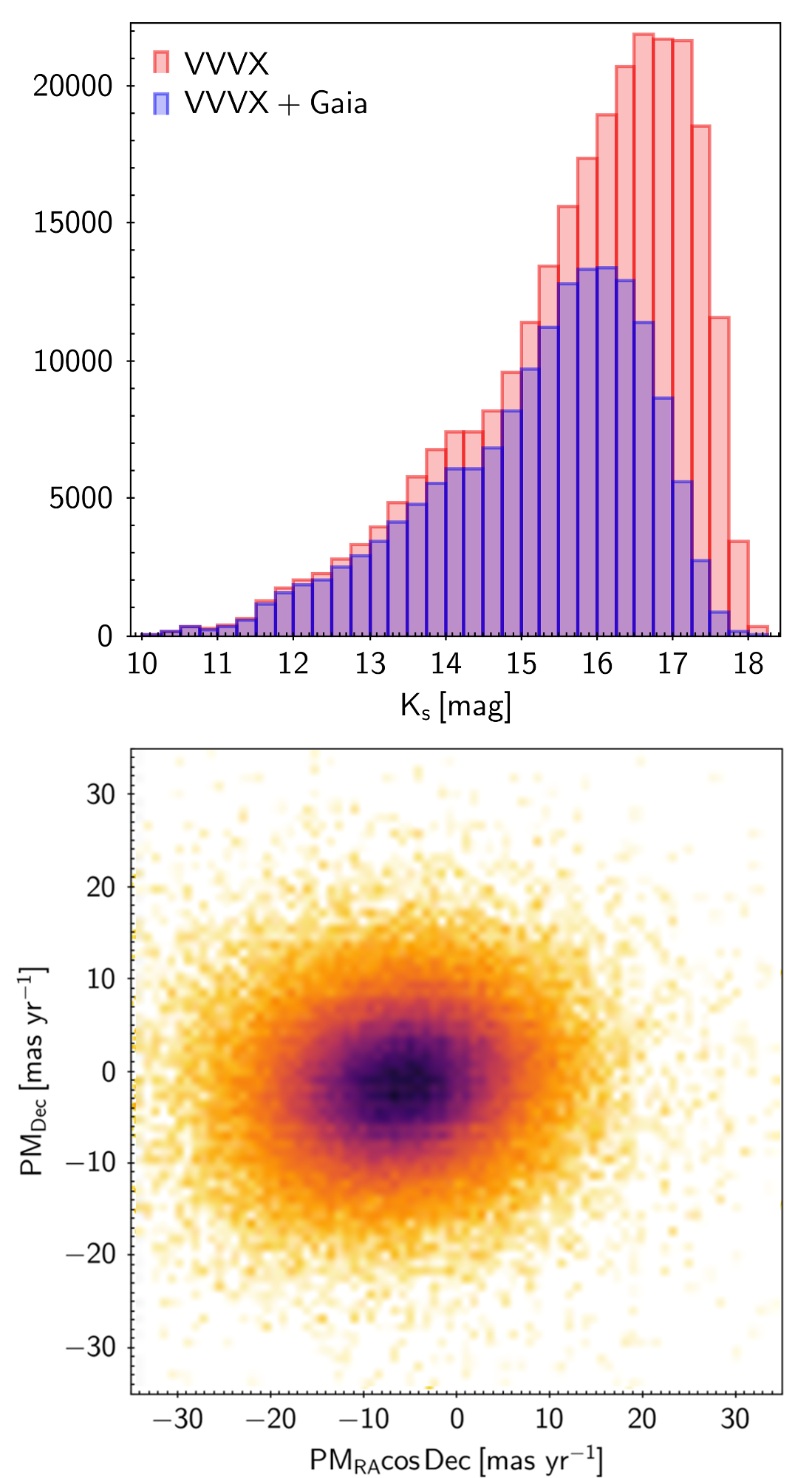}
\caption{Proper motion measurements for  tile e1084. Top: histogram in
  $K_{\rm s}$  band for stellar sources  in both the first  and latest
  $K_{\rm  s}$ band  epochs  observed  for e1084.  About  60\% of  the
  sources have counterparts in {\it Gaia} DR3.  This number dropped to
  $\lesssim$40\%  for  $K_{\rm  s}  > 16.0$  mag.   Bottom:  Naive  PM
  measurements  based on  the differences  of the  RA/DEC between  the
  latest and the first epochs.  The grid pattern is an artefact of the
  binning  algorithm used  for plotting,  and is  not a  real feature.
  Only stellar sources with $K_{\rm s}  < 16.0$ have been used and the
  timespan  is $\sim$\,4.2  years (1530  days).  The  mean values  are
  PM$_{\rm  RA}$\,cos\,Dec~$  =  -6.40  \pm  5.46$  mas~yr$^{-1}$  and
  PM$_{\rm Dec}= -1.53\pm5.02$ mas~yr$^{-1}$.}
\label{fig:pm}
\end{figure}

\subsection{Variability data: Light curves and proper motions}

\citet{2013ApJ...776L..19D}  present the  effects  of the  time-domain
sampling on the  detection efficiency of variable  stars.  They showed
that even a random cadence in the observations will allow us to detect
RR Lyrae and $\delta$-Cepheids  at $K_{\rm s}$=$14$~mag unambiguously,
provided a minimum number of  $25$ epochs.  Therefore, between $25-40$
individual  $K_{\rm s}$ band  observations were  allocated to  the new
extended  survey area.   More frequent  observations (40  epochs) were
assigned to more highly reddened regions  in the new VVVX disk area at
230$^{\circ}$$<$\,$l$\,$<$295$^{\circ}$  (see  Fig.~\ref{fig:epochs}),
given that fainter variable stars require a larger number of observing
epochs, and distant parts of the Galactic disk lie mainly in reddened,
low latitude regions.  Moreover, the  number of epochs also depends on
the  type of  variable star  we  expect to  find in  each region;  for
short-period  variables,  a  greater  number of  epochs  is  required.
Similarly, we  have simulated the  gain in proper motion  precision of
VVVX,  relative to  VVV.  We  find  that $10$~epochs  offers the  best
compromise between  precision and time spent,  yielding an uncertainty
of 300~$\mu$as/yr  thanks to  the 9.5 year  baseline.  This  very high
precision is essential to enable  high quality decontamination of star
clusters to $K_{\rm s}=15-16$ mag  in very crowded inner Galaxy fields
and our goal of 5D mapping  of Galactic structure.  VVVX astrometry in
VIRAC2  is placed  on the  {\it Gaia}  absolute astrometric  reference
frame, which is very  precise even in the plane due  to the {\it Gaia}
method of simultaneously observing two widely separated fields.

With  the  long  time-baseline,   the  high  precision  proper  motion
measurements  can disentangle  the  bulk  bulge/disk stellar  motions,
producing   a   global   pure  bulge/disk   colour-magnitude   diagram
\citep{2015MNRAS.450.1664L}. As the innermost  regions of the MW bulge
and central disk  remain out of reach for {\it  Gaia} observations due
to  their high  dust extinction,  it will  be up  to VVV  and VVVX  to
provide input catalogues for spectroscopic  surveys such as MOONS, the
Milky Way Mapper \citep[MWM; ][]{2017arXiv171103234K} and others.

The  observing strategy  for  e1084 consists  in  41 $K_{\rm  s}$ band
epochs over a  baseline of 4-years, which allows the  search and study
of  various  classes  of   variable  and  transient  stars,  including
pulsating variable stars used as distance indicators, such as RR Lyrae
and Cepheids. The  complete set of $K_{\rm  s}$ band source catalogues
for tile  e1084 were matched following  the same procedure as  for the
multicolour catalogue in  order to construct the light  curves for the
tile.   Fig.   \ref{fig:rrl}  shows   the  light  curves  derived  and
subsequently  phase folded  for two  previously known  RR Lyrae  stars
discovered   by   {\it   Gaia}   in   the   region   of   tile   e1084
\citep{2023A&A...674A..14R}.

By applying a Lomb  Scargle (LS) algorithm \citep{1976Ap&SS..39..447L,
  1982ApJ...263..835S, 2009A&A...496..577Z}  we detected  the periodic
signal, albeit  in the case  of Gaia DR3 5255393833932660992  with the
double period,  which is  a common  feature in  this type  of analysis
\citep[e.g.][]{2013BAAA...56..153C,2020MNRAS.496.1730F,
  2021MNRAS.504..654B}.   This  factor-of-two   error  in  the  period
determination is  a well-known issue  that affects the LS  method (and
its variants) when faced with  nearly sinusoidal light curves, both in
the near-IR  as in  the optical  \citep[e.g.][]{2013MNRAS.434.3423G,
  2017EPJWC.15203001G,  2018ApJS..236...16V}. In  applying the  string
length      minimisation       method      \citep{1970JRASC..64..353B,
  1983MNRAS.203..917D,  2002A&A...386..763C}, periods  of 0.4763  days
and 0.2380  days were  obtained for  Gaia DR3  5255393833932660992 and
Gaia DR3  5255378372049581568, respectively,  in agreement  with those
reported in the literature.

We emphasise that aperture data  is not ideal for studying variability
in  crowded  fields,   and  PSF  data  is  more   desirable,  or  even
differential imaging analysis (DIA), in  order to obtain more accurate
photometry and thus reduce the scattering on the light curves. We also
note that  there is a  potential problem  due to the  different epochs
having  slightly  different  zero  points,   so  it  is  necessary  to
recalibrate  the different  epochs to  a reference  zero point.   This
preliminary result using CASU aperture  data shows that VVVX data will
find variables even with a smaller number of epochs and a shorter time
baseline compared  to VVV,  especially in the  innermost disk  ($|b| <
1$~deg), where  surveys carried  out in the  optical, for  instance by
{\it Gaia} or  the upcoming {\it Vera C. Rubin}  Observatory with its
3200 megapixel camera, are highly affected by extinction.

As  described  above,  the  VIRAC catalogue  with  the  proper  motion
measurements based  on the PSF data  should be available soon  for the
VVVX area.  While VIRAC2 covers  the original VVV area  using VVV+VVVX
observations from  $2010-2019$, VIRAC3  will cover the  whole VVV+VVVX
area. VIRAC3 is currently being generated.

Regardless, we applied  a different method to the  CASU aperture data,
by subtracting  the differences of  the RA/DEC between the  latest and
the first epochs observed in $K_{\rm  s}$ for e1084 for $K_{\rm s}<16$
mag stellar sources (`$-1$' flag)  in both epochs, without any other
type of  calibration or selection  criteria.  Even with this  naive PM
approach, the  bulk motion is  in agreement with that  expected motion
for stars  in the Galactic disk  \citep[e.g.][]{2023A&A...674A..39G},
with <PM$_{\rm RA}$\,cos\,Dec>~$  = -6.40 \pm 5.46$  mas yr$^{-1}$ and
<PM$_{\rm   Dec}$>$=   -1.53\pm5.02$   mas   yr$^{-1}$.    (see   Fig.
\ref{fig:pm}). Given that {\it Gaia}  is severely limited in the inner
disk areas,  only about 40\%  of stellar  sources as faint  as $K_{\rm
  s}>16.0$   mag   are  detected,   compared   to   the  CASU   source
catalogues. These objects are likely the closest to us, that is, those
with bluer  colour in the  CMD.  Since {\it  Gaia} does not  reach the
stars closer to  the Galactic centre, VVVX and VVV  are invaluable for
PM studies  in those areas. This  is even more evident  when comparing
{\it Gaia} with PSF based source catalogues.

\section{The VVVX science returns}
\label{sec:returns}

A large  public survey such  as the VVVX is  expected to have  a broad
impact in  different areas of  astrophysics. As useful  references for
the reader interested  in using the VVVX database, in  this Section we
briefly summarise the published works on planetary, stellar, Galactic,
and  extragalactic   astronomy,  starting  with  those   that  include
large-scale photometry and astrometry.

\subsection{Photometry and astrometry}

As discussed  above, Alonso-Garc\'ia et  al.  (in prep.)   will extend
the near-IR  atlas from the VVV  footprint \citep{2018A&A...619A...4A}
to  the  VVVX  new  surveyed  regions in  the  Galactic  plane,  using
PSF-fitting techniques  to extract the  photometry. Smith et  al.  (in
prep.) will  also extend  the astrometric and  multi-epoch photometric
dataset from the  original VIRAC catalogue \citet{2018MNRAS.474.1826S}
using PSF photometry and adding the new epochs provided by the VVVX.

\subsection{Variable stars and Galactic astronomy}

Regarding  variable stars,  \citet{2021ASPC..529..199A} described  the
potential of  classical pulsators such  as RR Lyrae, Type  2 Cepheids,
and Classical Cepheids for tracing the structure of the inner Galactic
bulge    and    Disk   in    the    VVVX    region.    In    addition,
\citet{2023MNRAS.520..828D} performed the  automated classification of
eclipsing binary systems discovered in the extended database, that are
also useful tracers across the Galactic plane.

\citet{2022MNRAS.509.2566M}  performed  a  massive  classification  of
variable stars  across the Galactic bulge  and disk using the  VVV and
VVVX survey data, automatically classifying 1.4 million point sources.
In a  follow-up study,  \citet {2022MNRAS.517..257S} studied  the Mira
variable  stars  in the  innermost  regions  of  the MW,  using  these
luminous  variables as  tracers  of the  nuclear  stellar disk.   This
sample was  also used  to estimate  the epoch of  the Milky  Way's bar
formation        using       detailed        dynamical       modelling
\citep{2024MNRAS.tmp..735S}.

More  recently,  \citet{2024MNRAS.528.1789L}  carried  out  a  massive
search  for the  highest amplitude  infrared variables,  discovering a
number of eruptive protostars, a  new class of variables named dipping
giants, as well as a variety of other interesting sources (transients,
LPVs, microlensing events, etc.), thus enabling a variety of follow up
studies.    For  example,   \citet{2024MNRAS.528.1769G}  presented   a
detailed,  unique  multi-wavelength  study  of  an  ongoing  FUor-type
outburst. VVVX  will enable to  develop a unified picture  of eruptive
Young  Stellar  Objects  (YSOs)  by longer  term  monitoring  of  slow
variables  and  long-lasting YSO  eruptions,  to  bridge the  eruption
timescale gap between EXors (weeks-months) and FUors (>\,10~yr).

Large surveys also open the  door for serendipitous discoveries, which
often  lead to  new science  \citep[e.g.][]{2024arXiv240510427W}.  In
this  context, \citet{2023ApJ...958L...1S}  reported the  discovery of
VVV-WIT-12,  which appears  to  be a  four-year-long  period YSO  that
induces  variability  in  its  surrounding  nebula,  and  discuss  the
different possible scenarios that include a light echo or a precessing
circumstellar disk.

\subsection{Background galaxies}
\label{sub:background}

The VVVX images  are deep enough to see through  the MW, also enabling
studies of background galaxies in the ZoA. As an example, the Circinus
galaxy is the  nearest known Seyfert II galaxy, located  at a distance
$D  = 4$~Mpc.  \citet{2024MNRAS.tmp..724O}  carried out  a search  for
globular  clusters  (GCs)   in  the  halo  of   the  Circinus  galaxy,
discovering dozens of bonafide GC candidates.

More distant galaxies can  also be traced. \citet{2021MNRAS.502..601B}
and  \citet{2024A&A...682A.153D} discovered  thousands of  galaxies in
the ZoA behind the MW disk and bulge, respectively, that are useful to
complete  the   picture  of  large-scale  structure   in  the  region.
\citet{2023MNRAS.524..678D}  published  an  extensive  near-IR  galaxy
catalogue  in the  northern part  of the  Galactic disk  using machine
learning  techniques  for the  first  time  in these  regions.   Also,
\citet{2024A&A...686A..18M}    explored   the    classification   (and
misclassification)        of        galaxies       performed        by
\citet{2021MNRAS.503.5263Z}  using  machine   learning  tools  in  the
4XMM-DR9 database.

Clusters   of   galaxies  were   also   discovered   in  the   near-IR
database. \citet{2022A&A...663A.158G} published a deep near-IR view of
the Ophiuchus  galaxy cluster,  the second-brightest X-ray  cluster of
galaxies  in   the  sky,  after   the  Coma  cluster.    In  addition,
\citet{2023A&A...669A...7G} unveiled  a new such structure  behind the
MW.

\subsection{Low-mass stars and exoplanets}

The VVVX  near-IR database is also  useful for studies in  the area of
low luminosity stars and sub-stellar  objects (M dwarfs, white dwarfs,
brown   dwarfs,   and  giant   planets).   \citet{2022A&A...660A.131M}
presented deep  VVVX near-IR photometry  for 99 low-mass stars  in the
{\it  Gaia}  EDR3 Catalogue  of  Nearby  Stars,  very useful  for  the
characterisation of  individual objects and for  the identification of
new  faint objects  in our  vicinity.  C\'aceres et  al.  (2024,  {\it
  submitted})  reported  the  discovery   of  a  dozen  planetary-mass
binaries    in   the    Lower   Centaurus-Crux    association.   Also,
\citet{2024MNRAS.52710737F} presented  the study of a  benchmark White
Dwarf - Ultracool Dwarf wide field binary in the Galactic plane.

\subsection{Star clusters}

The VVVX depth and resolution permits  to see deep into our own Galaxy
\citep{2024A&A...683A.150M}. Therefore, the  database also enabled the
discovery  of  numerous star  clusters  in  the reddened  and  crowded
regions of the Galactic disk and bulge.

The   works    of   \citet{2018MNRAS.481.3902B,   2019BAAA...61..110B,
  2020MNRAS.499.3522B,    2021MNRAS.503.1864P,    2022MNRAS.513.5799P}
identified dozens  of open star  clusters, either new or  known, using
VVV(X) data,  enriching their  near-infrared cluster  sequences.  This
allowed the  authors to  characterise the  clusters at  different mass
ranges, probing the small scale star formation in the Galactic disk.

Numerous old  globular cluster were also  identified and characterised
in the VVVX areas of the  disk and bulge \citep[][Saroon et al.  2024,
  {\it      submitted}]{2018PASA...35...25B,      2021A&A...654A..39O,
  2019A&A...628A..45G,    2022MNRAS.509.4962G,    2020A&A...642L..19G,
  2021A&A...649A..86G,        2022A&A...658A.120G,2023A&A...669A.136G,
  2021ApJ...908L..42F}  and in  the Sagittarius  dwarf galaxy  located
behind the bulge \citep{2021A&A...654A..23G,2021A&A...650L..12M}.  The
near-IR photometry allows the  determination of some important cluster
parameters,  such  as  reddening, distance,  luminosity,  metallicity,
mass, structure,  and age,  contributing to  the understanding  of the
Milky  Way GC  system as  a whole  \citep[][Garro et  al.  2024,  {\it
    submitted}]{2021BAAA...62..107M}.

\subsection{Long timescale microlensing events}
\label{sub:micolensing}

VVV  has  demonstrated  its  ability to  detect  microlensing  events,
particularly in highly obscured and  crowded fields, where such events
are  more  frequent  \citep{2017ApJ...851L..13N,  2018ApJ...865L...5N,
  2020ApJ...893...65N,       2020ApJ...889...56N}.       Long-duration
microlensing events  exceeding one  year are excellent  candidates for
stellar  mass  black  holes.   The  extended  time  baseline  of  VVVX
observations  will enable  the  discovery of  such  events around  the
Galactic  centre  and plane,  where  models  predict the  presence  of
stellar   mass    black   holes   \citep[e.g.][]{1976ApJ...209..214B,
  2006ApJ...649...91F, 2009ApJ...697.1861A}.

\section{Survey legacy}
\label{sec:legacy}

The VVV  and VVVX surveys  are the result of  more than 4000  hours of
observation    with   the    most   advanced    ground-based   near-IR
facility. Although VVV+VVVX had  observed $\sim\,$4\% of the celestial
sphere, the region contains the majority  of the Milky Way's stars, as
well as the largest concentration of gas and dust in the Galaxy.

There   are  currently   no  other   near-IR  projects   with  similar
characteristics to VVV+VVVX, such  as time-baseline, wavelength range,
photometric  depth, and  most importantly,  the large  projected area.
However, the legacy of both  surveys is much enhanced by complementary
data from DECaPS and Pan-STARRS \citep{2016arXiv161205560C}, that also
cover the MW bulge and southern  plane.  Here we must also mention the
complementary survey VISIONS  \citep{2023A&A...673A..58M} that targets
individual star forming regions in the MW.
 
In the  future projects, such as  the {\it Vera C.  Rubin} Observatory
which will  use optical wavelengths  for massive variability,  it will
also  be   very  complementary.   Also  complementary   are  the  JWST
observations, with instruments covering a small field of view but with
much higher resolution and depth (e.g., Section 7).  However, it would
be  unfeasible  for the  JWST  to  cover  large  areas with  the  same
efficiency as our survey.

Clearly, the VVVX  survey also serves as a vast  source of targets for
spectroscopic  follow-up with  future  infrared spectroscopic  surveys
such  as  MOONS,  4MOST,  MWM and  the  ESO  Wide-field  Spectroscopic
Telescope \citep[WST;][]{2024arXiv240305398M}. Our survey is  also a
source  of targets  for  the various  next-generation extremely  large
telescopes;  and may  only be  surpassed  in performance  in the  next
decade  by  projects  such  as  the  {\it  Nancy  Grace  Roman}  Space
Telescope, which will be able to produce deep infrared images of large
regions, including  the Galactic  centre, with higher  resolution from
space \citep[e.g.][]{2023arXiv230707642P}.

The VVVX  (and VVV) legacy will  last for many years  to come. Despite
the numerous  results already obtained,  the full exploitation  of the
data will take many years more, becoming more attractive with each new
product  that  can  be  obtained,  as  detailed  extinction  (or  even
metallicity)  maps,  catalogues  of  variable  sources  and  transient
objects, and  even studies of  background galaxies, especially  in the
ZoA beyond the MW plane.

\section{Summary}
\label{summary}

In March  2023 we finished  successfully the VVVX  survey observations
that  started in  2016, which  represents a  huge amount  of data  and
processing,  comprising $\sim  200,000$  images  that monitor  $>10^9$
sources in our Galaxy and beyond.  This is quite an accomplishment for
VISTA,  for ESO's  Paranal Observatory,  for the  CASU data  reduction
pipeline, for the teams extracting PSF photometry from the images, for
the VISTA  Science archive in Edinburgh  and for the ESO  archive: the
successful  VVVX  observations  are  now  100\%  completed.   We  have
described the  survey, including the observations,  areal and temporal
coverage,  reductions,  photometry,  and   astrometry.  We  have  also
presented  some  specific  scientific  examples  for  database  usage,
providing some key useful references.

Clearly, there  are many more  applications of this ESO  Public Survey
for the community to exploit for future studies of Galactic structure,
stellar populations, variable stars, star  clusters of all ages, among
other  exciting  research  areas,   from  stellar  and  (exo)planetary
astrophysics  to extragalactic  studies.  The  image processing,  data
analysis and  scientific exploitation will  continue for the  next few
years, with many discoveries yet to come. The VVVX Survey will also be
combined with  future facilities  to boost  its scientific  outcome in
unpredictable  ways:  we are  sure  that  this  survey will  remain  a
goldmine for MW studies for a long time.

\begin{acknowledgements}

We gratefully acknowledge  the use of data from the  ESO Public Survey
program IDs  179.B-2002 and 198.B2004  taken with the  VISTA telescope
and data products  from the Cambridge Astronomical  Survey Unit (CASU)
and the VISTA  Science Archive (VSA) and the ESO  Science Archive. VVV
and VVVX  data are published  in the ESO  Science Archive in  the data
collections      identified      by      the      following      DOIs:
{\tt\tiny{https://doi.eso.org/10.18727/archive/67}}                and
{\tt\tiny{https://doi.eso.org/10.18727/archive/68}}.            R.K.S.
acknowledges support from  CNPq/Brazil through projects 308298/2022-5,
350104/2022-0  and   421034/2023-8.   D.M.    gratefully  acknowledges
support from  the Center for Astrophysics  and Associated Technologies
CATA by  the ANID BASAL  projects ACE210002 and FB210003,  by Fondecyt
Project   No.    1220724,   and   by   CNPq/Brazil   through   project
350104/2022-0.   J.A.-G., acknowledges  support from  Fondecyt Regular
1201490  and  by ANID  --  Millennium  Science Initiative  Program  --
ICN12\_009 awarded  to the  Millennium Institute of  Astrophysics MAS.
Support  for J.B.  and R.K.  are provided  by ANID's  FONDECYT Regular
grant \#1240249;  ANID's Millennium Science Initiative  through grants
ICN12\_009  and AIM23-0001,  awarded  to the  Millennium Institute  of
Astrophysics (MAS).  C.C.  acknowledges  support by ANID BASAL project
FB210003.  N.J.G.C.   acknowledges support  from  the  UK Science  and
Technology Facilities Council. E.B.A.  thanks Universidade Estadual de
Feira  de Santana  for the  support received  by the  Program FINAPESQ
(project number  050/2021). L.R.B.  acknowledges financial  support by
INAF under WFAP project, f.o.:1.05.23.05.05.  J.I.A.  acknowledges the
financial support of DIDULS/ULS,  through the project PR2324063.  A.B.
acknowledges  support from  the Deutsche  Forschungsgemeinschaft (DFG,
German Research Foundation) under  Germany's Excellence Strategy - EXC
2094  -  390783311.   B.D.    acknowledges  support  by  ANID-FONDECYT
iniciaci\'on  grant  No. 11221366  and  from  the ANID  Basal  project
FB210003.  A.C.G.  acknowledges support  from PRIN-MUR 2022 20228JPA3A
``The  path to  star and  planet formation  in the  JWST era  (PATH)''
funded by NextGeneration EU and  by INAF-GoG 2022 ``NIR-dark Accretion
Outbursts in Massive  Young stellar objects (NAOMY)''  and Large Grant
INAF  2022 ``YSOs  Outflows,  Disks and  Accretion:  towards a  global
framework  for  the  evolution  of planet  forming  systems  (YODA)''.
J.A.C-B.   acknowledges  support  from  FONDECYT  Regular  N  1220083.
Support  for  M.C.   is  provided by  ANID's  FONDECYT  Regular  grant
\#1171273;  ANID's   Millennium  Science  Initiative   through  grants
ICN12\textunderscore  009 and  AIM23-0001, awarded  to the  Millennium
Institute of  Astrophysics (MAS);  and ANID's Basal  project FB210003.
M.C.C.    acknowledges   financial   support  from   the   Universidad
Complutense de Madrid (UCM) and the Agencia Estatal de Investigaci\'on
(AEI/10.13039/501100011033)   of   the   Ministerio   de   Ciencia   e
Innovaci\'on and the ERDF ``A  way of making Europe'' through projects
PID2019-109522GB-C5[4]  and  PID2022-137241NBC4[4].   P.C.   and  E.S.
acknowledge  financial support  from the  Spanish Virtual  Observatory
project funded by the Spanish Ministry of Science and Innovation/State
Agency  of   Research  MCIN/AEI/10.13039/501100011033   through  grant
PID2020-112949GB-I00. V. Motta acknowledges support from ANID FONDECYT
Regular  grant number  1231418.  J.G.F-T  gratefully acknowledges  the
grants support  provided by ANID Fondecyt  Iniciaci\'on No.  11220340,
ANID Fondecyt  Postdoc No.  3230001 (Sponsoring  researcher), from the
Joint Committee ESO-Government  of Chile under the  agreement 2021 ORP
023/2021 and 2023 ORP 062/2023.   Support for C.E.F.L.  is provided by
the   ANID/FONDECYT   Regular   grant   1231637.    D.G.    gratefully
acknowledges  the support  provided by  Fondecyt regular  n.  1220264.
D.G.   also acknowledges  financial  support from  the Direcci\'on  de
Investigaci\'on y  Desarrollo de la  Universidad de La  Serena through
the  Programa  de  Incentivo  a  la  Investigaci\'on  de  Acad\'emicos
(PIA-DIDULS).  The work  of F.  N.  is supported by  NOIRLab, which is
managed by the  Association of Universities for  Research in Astronomy
(AURA)  under  a  cooperative  agreement  with  the  National  Science
Foundation.  W.G.   gratefully acknowledges funding from  the European
Research  Council  (ERC)  under  the  European  Union's  Horizon  2020
research  and  innovation  programme   under  grant  agreement  951549
(project UniverScale).  F.G.  gratefully acknowledges support from the
French  National Research  Agency (ANR)  - funded  projects ``MWDisc''
(ANR-20-CE31-0004)  and  ``Pristine''  (ANR-18-CE31-0017).   Z.G.   is
supported  by the  ANID  FONDECYT Postdoctoral  program No.   3220029.
E.L.M.  is supported  by the European Research  Council Advanced grant
SUBSTELLAR, project number 101054354.  V.M.  acknowledges support from
project DIDULS Regular N$^{o}$ PR2353857.  M.C.  thanks the support of
ANID  BECAS/DOCTORADO NACIONAL  21110001.  I.P.   acknowledges support
from  ANID  BECAS/DOCTORADO  NACIONAL  21230761.   G.P.   acknowledges
support from ANID through Millennium Science Initiative Programs ICN12
009.  S.R.A.  acknowledges support from Fondecyt Regular 1201490. E.S.
acknowledges financial  support from  the Spanish  Virtual Observatory
project funded by the Spanish Ministry of Science and Innovation/State
Agency  of   Research  MCIN/AEI/10.13039/501100011033   through  grant
PID2020-112949GB-I00.  D.S.  acknowledged support from the Science and
Technology  Facilities  Council  (STFC,  grant  numbers  ST/T007184/1,
ST/T003103/1, ST/T000406/1  and ST/X001121/1).  M.T.  is  supported by
the  JSPS  Kakenhi  No.   24H00242.   P.B.T.   gratefully  acknowledge
support by the ANID BASAL project FB210003 and Fondecyt 1240465.  S.V.
gratefully  acknowledges the  support provided  by Fondcyt  regular n.
1220264.   SV  gratefully  acknowledges  support  by  the  ANID  BASAL
projects  ACE210002  and  FB210003.  C.N.M.   gratefully  acknowledges
support  from the  Research Department  of the  Austral University  of
Chile, Puerto Montt Campus  (Project DIPM-CIB2303).  Financial support
for  this  work was  also  provided  by  the]  ANID BASAL  Center  for
Astrophysics  and   Associated  Technologies  (CATA)   through  grants
AFB170002, ACE210002 and FB210003, by the ANID Millennium Institute of
Astrophysics  (MAS) ICN12\_009  and by  ANID Fondecyt  Regular 1230731
(PI: M.Z.).  \\  \textbf{ \textit{ We are also deeply  thankful to our
    dear colleague  Rodolfo Barb\'a, who  was a pillar for  the survey
    but sadly passed away in late 2021, may he rest in peace.}}

\end{acknowledgements}

\begin{appendix}
\section{VVVX tile coordinates and observations}
\label{app:tiles}

Here  we  list the  tile  centre  coordinates  for  all VVV  and  VVVX
pointing.  There  is a total of  1028 tiles, divided into  348 for the
original  VVV area  and  680 tiles  for VVVX.   For  the original  and
extended bulge area tiles, names start with ``b''. Inner disk tiles in
the original and extended area start with ``d'', while for the low and
high disk,  as well  as to  disk to longitude  $+$20 names  start with
``e''.   Fig.  \ref{fig:vvvx}  shows the  survey area  with the  tiles
positions and respective  names. For each tile we  provide tile centre
coordinates in Equatorial  and Galactic systems.  All  tiles have been
observed  using  an  identical   offsetting  strategy,  combining  six
pawprints  to  contiguously  fill  $1.5  \times  1.1$~sq.~deg.   area.
Columns 6, 7 and 8 present the  number of epochs taken in $J$, $H$ and
$K_{\rm s}$ during the VVV and VVVX campaigns. The first number is the
total of epochs,  and in parentheses the number of  epochs observed in
VVV and VVVX, respectively.

\begin{figure*}
\begin{centering}
\includegraphics[angle=90,scale=0.96]{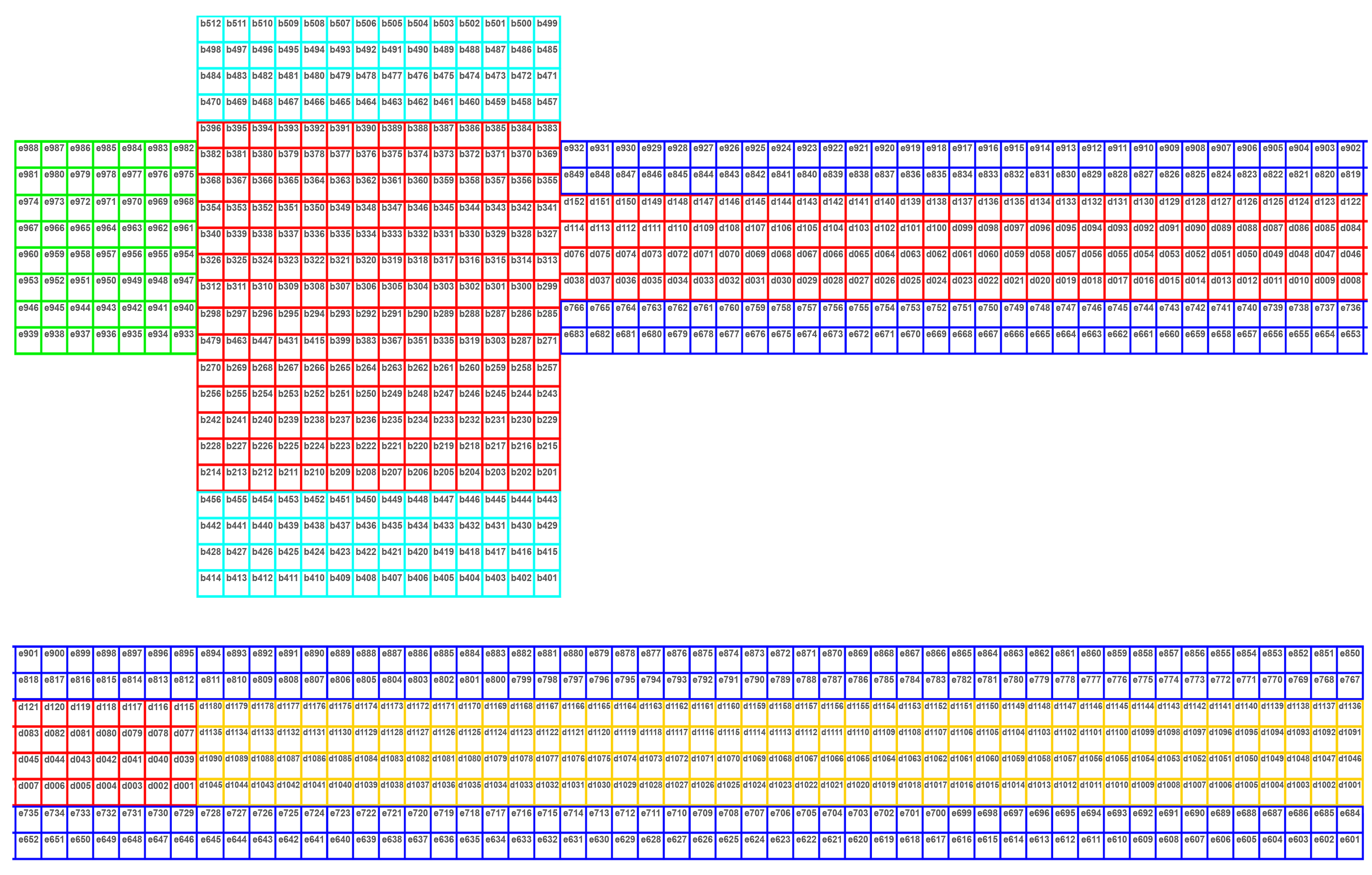}
\caption{Zoomed view of the VVVX  survey area with the tiles positions
  and respective names.  The original VVV area in shown  in red, while
  the  other  colours mark  the  VVVX  areas.   For the  original  and
  extended bulge area tile, names  start with `b'.  Inner disk tiles
  in the  original and extended area  start with `d', while  for the
  low and  high disk, as well  to disk to longitude  $+$20 names start
  with `e'.}
\label{fig:vvvx}
\end{centering}
\end{figure*}

\clearpage
\onecolumn

\begin{longtable}{lccrrrrr}
\caption{\label{tab:tiles} List the VVV+VVVX tiles and observed number
  of  epochs. We  present the  tile  names and  centre coordinates  in
  Equatorial and  Galactic systems  (see Fig.  \ref{fig:vvvx}  for the
  spatial distribution).  Columns 6 to  8 present the number of epochs
  observed  in $J$,  $H$  and  $K_{\rm s}$  during  the  VVV and  VVVX
  campaigns.   The  first  number  is  the total  of  epochs,  and  in
  parentheses  the  number  of  epochs   observed  in  VVV  and  VVVX,
  respectively.}\\ \hline\hline Tile & RA  (J2000.0) & DEC (J2000.0) &
longitude &  latitude & $J$ epochs  & $H$ epochs &  $K_{\rm s}$ epochs
\\ name &  (dd:mm:ss.ss) & (dd:mm:ss.s) & (degrees) &  (degrees) & & &
\\ \hline \endfirsthead
\caption{continued.} \\
\hline\hline
Tile & RA (J2000.0)    & DEC (J2000.0) & longitude & latitude  & $J$ epochs & $H$ epochs & $K_{\rm s}$ epochs \\
name & (hh:mm:ss.ss)   & (dd:mm:ss.s)  & (degrees) & (degrees) &            &            &                  \\
\hline
\endhead
\hline
\endfoot
b0201 & 18:04:36.67 & $-$41:52:19.1 &   $-$9.344772 &  $-$9.781310 &  $ 2~~( 2+ 0)$ &  $ 2~~( 2+0)$ & $   79~~( 76 +   3)$ \\
b0202 & 18:08:12.42 & $-$40:34:54.6 &   $-$7.866743 &  $-$9.781265 &  $ 2~~( 2+ 0)$ &  $ 2~~( 2+0)$ & $   80~~( 76 +   4)$ \\
b0203 & 18:11:41.75 & $-$39:17:16.9 &   $-$6.388843 &  $-$9.781293 &  $ 2~~( 2+ 0)$ &  $ 2~~( 2+0)$ & $   81~~( 77 +   4)$ \\
b0204 & 18:15:05.21 & $-$37:59:26.7 &   $-$4.910874 &  $-$9.781311 &  $ 2~~( 2+ 0)$ &  $ 2~~( 2+0)$ & $   78~~( 75 +   3)$ \\
b0205 & 18:18:23.31 & $-$36:41:25.3 &   $-$3.432812 &  $-$9.781345 &  $ 2~~( 2+ 0)$ &  $ 2~~( 2+0)$ & $   78~~( 75 +   3)$ \\
b0206 & 18:21:36.47 & $-$35:23:14.3 &   $-$1.954806 &  $-$9.781319 &  $ 2~~( 2+ 0)$ &  $ 2~~( 2+0)$ & $   83~~( 79 +   4)$ \\
b0207 & 18:24:45.14 & $-$34:04:55.1 &   $-$0.476882 &  $-$9.781341 &  $ 2~~( 2+ 0)$ &  $ 2~~( 2+0)$ & $   82~~( 79 +   3)$ \\
b0208 & 18:27:49.70 & $-$32:46:28.3 &      1.001059 &  $-$9.781314 &  $ 2~~( 2+ 0)$ &  $ 2~~( 2+0)$ & $   80~~( 76 +   4)$ \\
\end{longtable}

\clearpage
\twocolumn

\end{appendix}

\end{document}